\documentclass[english,aps,preprint,superscriptaddress,longbibliography]{revtex4-1}

\usepackage{graphicx}
\usepackage[colorlinks=true,allcolors=blue]{hyperref}
\usepackage{multirow}
\usepackage{longtable}
\usepackage{rotating}
\usepackage{rotate}

\usepackage{lineno,xcolor}
\usepackage{amsmath,amssymb}
\usepackage{array}

\begin{document}
\title{Quantitatively ranking incorrect responses to multiple-choice questions using item response theory}
\author{Trevor I.\ Smith}
\affiliation{Department of Physics and Astronomy, Rowan University, Glassboro, NJ 08028, USA}
\affiliation{Department of STEAM Education, Rowan University, Glassboro, NJ 08028, USA}

\author{Kyle J.\ Louis}
\affiliation{Department of Physics and Astronomy, Rowan University, Glassboro, NJ 08028, USA}
\affiliation{Department of STEAM Education, Rowan University, Glassboro, NJ 08028, USA}

\author{Bartholomew J.\ Ricci, IV}
\affiliation{Department of Physics and Astronomy, Rowan University, Glassboro, NJ 08028, USA}
\affiliation{Department of Mathematics, Rowan University, Glassboro, NJ 08028, USA}

\author{Nasrine Bendjilali}
\affiliation{Department of Mathematics, Rowan University, Glassboro, NJ 08028, USA}

\begin{abstract}
Research-based assessment instruments (RBAIs) are ubiquitous throughout both physics instruction and physics education research. The vast majority of analyses involving student responses to RBAI questions have focused on whether or not a student selects correct answers and using correctness to measure growth. This approach often undervalues the rich information that may be obtained by examining students' particular choices of incorrect answers. In the present study, we aim to reveal some of this valuable information by quantitatively determining the relative correctness of various incorrect responses. To accomplish this, we propose an assumption that allow us to define relative correctness: students who have a high understanding of Newtonian physics are likely to answer more questions correctly and also more likely to choose better incorrect responses, than students who have a low understanding. Analyses using item response theory align with this assumption, and Bock's nominal response model allows us to uniquely rank each incorrect response. We present results from over 7,000 students' responses to the Force and Motion Conceptual Evaluation.
\end{abstract}

\maketitle

\section{Introduction}
\label{sec:intro}
Many instructional and research questions over the past three decades have been answered by examining student responses to multiple-choice Research-based Assessment Instruments (RBAIs) \cite{Madsen2017,VonKorff2016}. Tens of thousands of students have provided responses to questions on the Force Concept Inventory (FCI \cite{Hestenes1992}), the Force and Motion Conceptual Evaluation (FMCE \cite{Thornton1998}), and many others \cite{physport}, and dozens of analyses have been published that use these results to measure student learning (See Refs.\ \cite{Docktor2014} and \cite{Madsen2017}). A common factor throughout most of these analyses is that students' responses are typically scored as being correct or incorrect; very little attention has been paid to which incorrect answers students choose. This dichotomous scoring scheme is very beneficial for simplifying student performance on a RBAI or growth in learning to a single number that may be compared between students or across populations. The simplicity of this analysis and the ability for instructors and researchers to compare their results with other data sets has contributed to the proliferation of RBAIs, to the benefit of the physics education research (PER) community; however, the dichotomous scoring scheme implicitly ignores any information about students' choices that are not correct. All incorrect answers are treated equally, regardless of how similar or different they may be to the correct answer.

RBAIs are so powerful because their questions help to elicit students' core beliefs about how the world works in ways that mathematical or problem-solving questions often do not. Many of the incorrect ``distractor'' response choices correspond with deeply held intuitive understandings that fit well with everyday experiences (and correspond with historically accurate models) but conflict with the principles of Newtonian physics \cite{Madsen2017}. The ability to deeply probe students' conceptual understanding of physics and represent this understanding with a single numerical value is very powerful. The authors of the FMCE, in fact, argue against using a single numerical score to represent student understanding \cite{Thornton2009} instead favoring the examination of student performance on individual or small groups of questions \cite{Thornton1998}, but the common practice persists. Moreover, the common practice of reporting normalized gain as a measure of student learning has been shown to be biased against students with little prior exposure to formal physics instruction \cite{Nissen2018a}. 

Other analyses of student responses to RBAI questions examine specific choices that students make and relate these choices to various mental models \cite{Bao2006}, misconceptions \cite{Hestenes1992}, views \cite{Thornton1997}, or pieces of knowledge \cite{Smith2008,Smith2014} that students may have or use when answering particular questions. These analyses provide a lot of rich information about students' ideas, but the processes of conducting these analyses are often quite time intensive, and the presentation and visualization of the results can be conceptually dense and difficult to interpret \cite{Smith2015c,Griffin2016}. As such, these analyses are not nearly as common as reporting a single numeric score.

Our ultimate goal is to define a single numeric score that represents student knowledge or understanding as measured by a RBAI by incorporating both correct responses and the good ideas that may be expressed in some incorrect responses. The first part of defining such a metric is to determine whether or not some incorrect answers may be considered better than others, where ``better'' means closer to correct or indicating a higher level of understanding. Considering one incorrect answer to be better than another can be a tricky business, and we want to make sure that we are not introducing personal bias into our definitions. As such, we carefully articulate the assumption for defining what makes one response better than another, and we choose an analysis method that correspond to that assumption to quantitatively rank incorrect responses based on students' response patterns.

\begin{quote}
\textbf{Assumption: Students who choose correct responses on most questions are more likely to choose better incorrect answers than students who choose few correct responses.} 
\end{quote}

This assumption is based on the premise that students who understand more about Newtonian physics are more likely to choose better incorrect answers than students who understand less physics, and these students are also more likely to choose a greater number of correct responses. This assumption is consistent with previous work that has used item response curves (IRCs) to examine and rank incorrect responses on both the FCI and the FMCE \cite{Morris2012,Walter2016,Ishimoto2017,Smith2017}. We expand on this prior work by using a nested-logit item response theory (IRT) model to simultaneously estimate students' overall understanding of Newtonian mechanics (the IRT latent trait, or person parameter) and determine how closely each response choice correlates with a high level of understanding using the estimated parameters of the model \cite{Bock1972,Thissen2010,Suh2010,Chalmers2012,mirt,Louis2018}. Based on this assumption we would claim, for example, that a student who only incorrectly answers one question is more likely to choose a response that's almost correct than a student who answers 20 questions incorrectly.

To illustrate the applicability of our assumption, we analyzed more than 7,000 students' matched pre-/post-test responses to the FMCE to demonstrate how quantitative analyses can provide information about which response choices may be better than others. We present a ranking of incorrect responses for all FMCE questions as well as the parameter values used to make these determinations.

\section{Data Sources and Preparation}
\label{sec:data}
Our data come from two primary sources:
\begin{itemize}
    \item sets of student responses to the FMCE provided to one author (TIS) by colleagues from four different colleges or universities, Schools 1--4,\footnote{Analyses of data from School 1--2 was published in Ref.\ \cite{Smith2014}.} as part of current and previous research projects ($N=952$), and
    \item student responses uploaded to PhysPort's Data Explorer ($N=6,336$) \cite{physportde}.
\end{itemize}
Some information is known about the instructional settings at Schools 1--4 (all of which used research-based instructional materials of some sort), but this information is not available for the PhysPort data. For the purposes of the current analysis, we combine all data into one set of $N=7,288$ students. We are not interested in how instructional factors impact student learning for this analysis, or whether or not student responses are different before or after instruction. As such, we have combined all pretest and posttest responses into a data set of $N=14,576$ response sets.

To prepare the data for analysis, we omit any responses that are inappropriate for a given question (e.g., a given response of E on question 45, which only includes options A, B, C, and D). We also omit response J (None of these answers is correct) from interpretations of our analyses because it does not represent a well-defined indication of what each student would consider correct: two students who choose answer A agree on what they consider to be correct, but two students who choose answer J may have very different ideas of what would be a correct answer, so we cannot claim similarities between the responses of students who choose J. We also removed response sets with three or more blank or unscorable responses. This gave us a usable data set of $N=12,388$ response sets.

The structure of the FMCE makes it an interesting focus for this work. Unlike many other RBAIs, the FMCE contains several questions for each physical scenario presented (e.g., a toy car moving horizontally), and all questions in each set have the same set of response choices. This is particularly interesting because a response choice that corresponds with the most common intuitive answer to one question, may not relate to any documented reasoning for another question.

\section{The 2PL-NRM Nested Logit Model}
\label{sec:irt}
Item Response Theory (IRT) uses students' responses to multiple-choice questions to simultaneously estimate each student's overall understanding of the material (a.k.a.\ the latent trait or person parameter, $\theta$) and determine the probability that a student will be correct on each question given his/her understanding \cite{Baker2001,deAyala2008}. The latent trait is normalized such that the average value is $\langle\theta\rangle=0$ and the standard deviation is $\sigma_\theta=1$. In the two-parameter logistic (2PL) IRT model, the probability of a student answering a specific question correctly is given by,
	\begin{align}
		P\left(\theta\right)&=\frac{1}{1+e^{-a\left(\theta-b\right)}}
		\label{2pl}
	\end{align}
where $a$ is the discrimination parameter and $b$ is the difficulty parameter.  Some previous work has used the three-parameter logistic model to analyze RBAI data \cite{Wang2010}, but we feel that the inclusion of the third ``guessing'' parameter is inappropriate for our analyses given that student responses to the FMCE are concentrated in a small subset of responses for each question: they are not, in fact, guessing \cite{Thornton2009,Smith2008}. 

\begin{figure}
    \centering
    \begin{tabular}{cc}
    \includegraphics[width = 0.225 \textwidth]{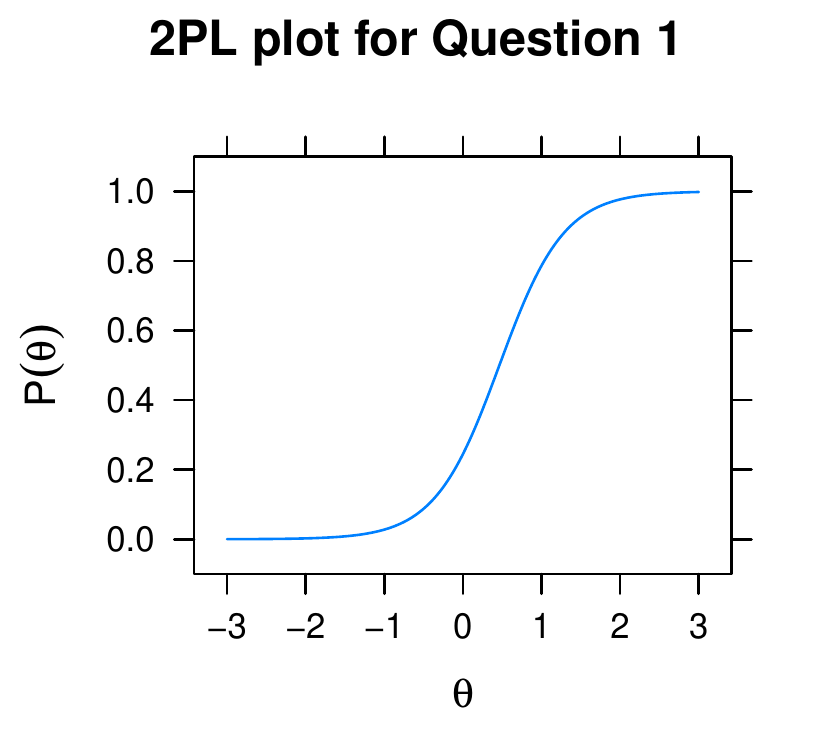}& 
    \includegraphics[width = 0.225 \textwidth]{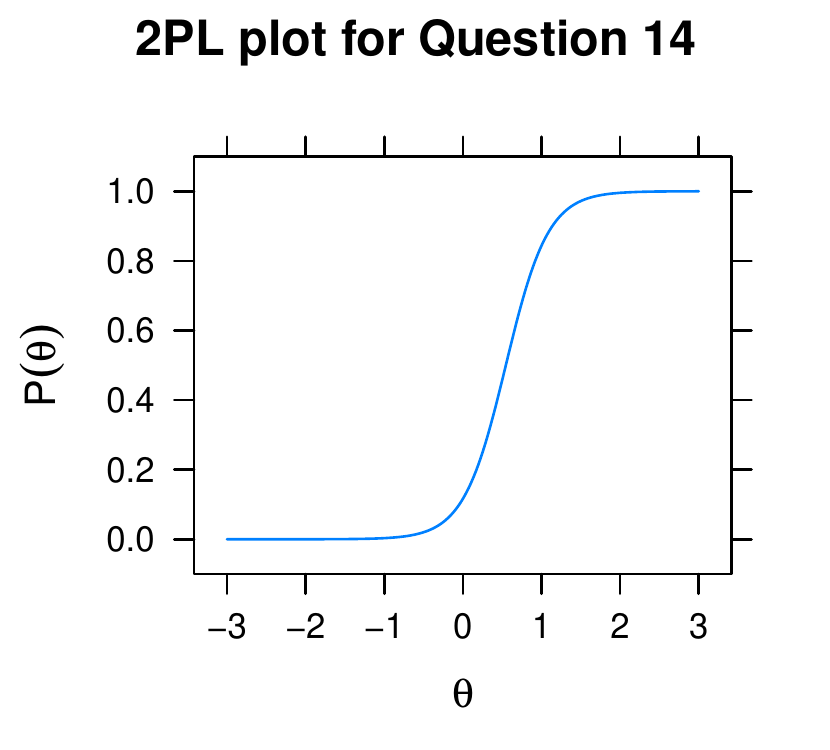}\\
    (a) Question 1 & (b) Question 14\\ 
    $a=2.5$, $b=0.45$ &$a=3.7$, $b=0.54$\\[5ex]
    \includegraphics[width = 0.225 \textwidth]{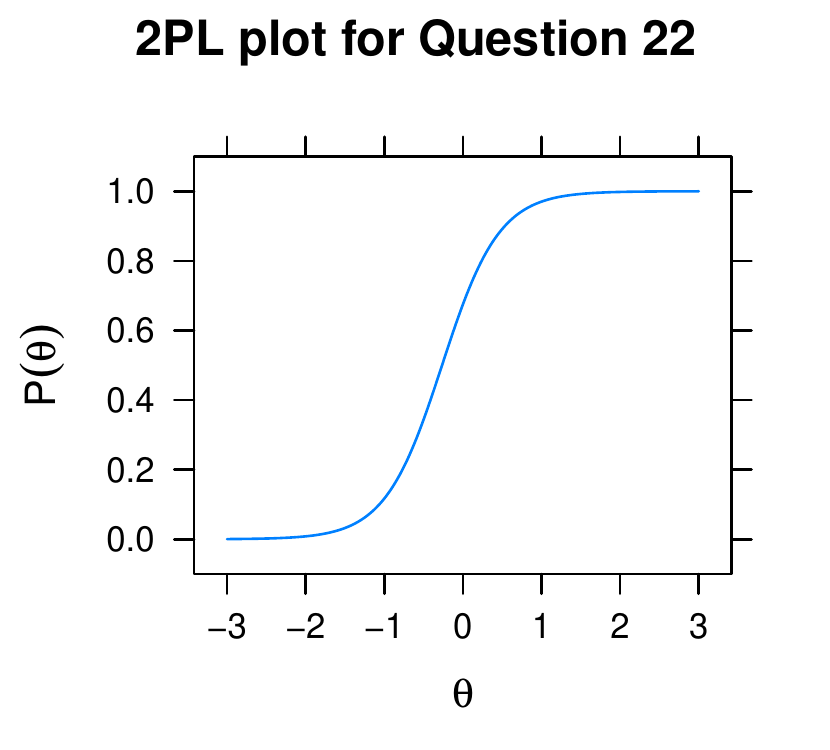} &
    \includegraphics[width = 0.225 \textwidth]{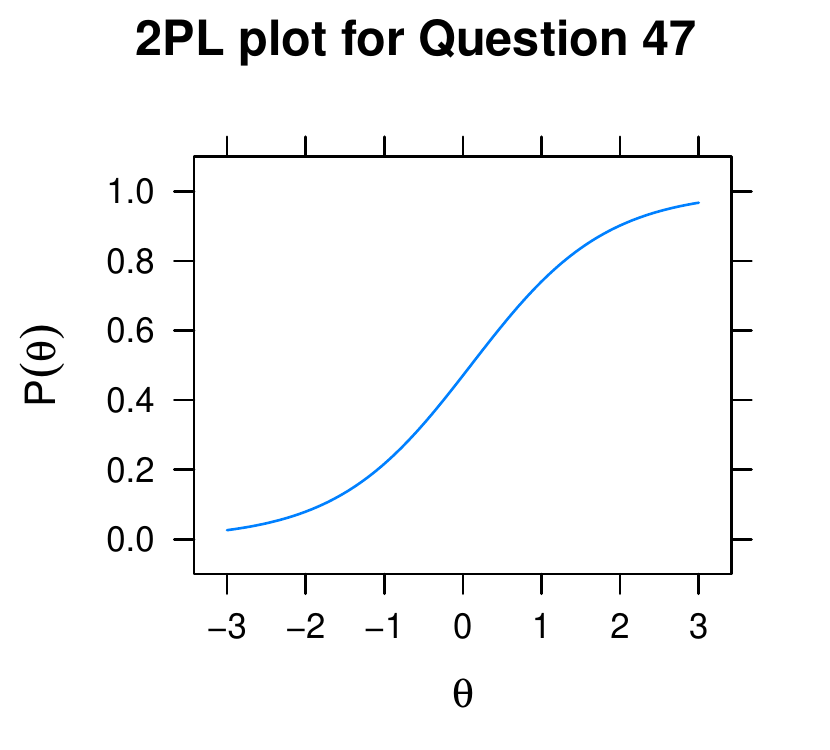}\\
    (c) Question 22 & (d) Question 47\\
    $a=2.8$, $b=-0.27$ & $a=1.1$, $b=0.10$
    \end{tabular}

    \caption{Item response theory (IRT) plots for questions 1, 14, 22, and 47: probability of answering correctly as a function of the latent ability parameter $\theta$. The plots help to illustrate the meaning of the discrimination $a$ and difficulty $b$ parameters in the two-parameter logistic (2PL) IRT model shown in Eq.\ \eqref{2pl}.}
    \label{2plPlots}
\end{figure}

The interpretation of the parameters in the 2PL model may be understood by examining plots of $P(\theta)$ vs.\ $\theta$: Fig.\ \ref{2plPlots} shows examples from several questions. The difficulty $b$ is the value of $\theta$ at which $P(b) = 0.5)$, and the discrimination $a$ is proportional to the slope of the curve at $\theta = b$: $\mathrm{d}P/\mathrm{d}\theta|_{b} = a/4$. Questions 1 and 14 (Fig.\ \ref{2plPlots} (a) and (b), respectively) have similar difficulty parameters (the $b$ value differs by less than 0.1 standard deviations of the latent ability $\theta$), but Q14's higher discrimination parameter $a$ shows up as a sharper transition from most likely incorrect to most likely correct, and a steeper slope at the midpoint of the curve. Question 22 in Fig.\ \ref{2plPlots}(c) has a similar discrimination to Q1 (similar slope at $P(\theta)=0.5$), but the difficulty is much lower (shown by a shift to the left compared to Fig.\ \ref{2plPlots}(a)), with many below-average students ($\theta<0$) being fairly likely to answer correctly. Question 47 in Fig.\ \ref{2plPlots}(d) has a difficulty parameter that is about average (close to zero), but the discrimination is relatively small, as shown by a shallow slope, and a more gradual transition from probably incorrect to probably correct than any of the other three. Higher values of discrimination $a$ mean a sharp transition and steeper slope; lower values mean a gradual transition and shallower slope. Higher values of difficulty $b$ mean a graph that is shifted to the right; lower values mean a graph that is shifted to the left. 

Bock's nominal response model (NRM) provides the probability of a person choosing each possible response based on $\theta$ \cite{Bock1972},
\begin{align} 
	P_k(\theta)&= \frac{e^{a_k\left(\theta-b_k\right)}}{\sum\limits_{i=1}^{N} e^{a_i\left(\theta-b_i\right)}},
	\label{nrm}
\end{align}
where $k$ indicates the particular response choice, and the summation is performed over all $N$ response choices. According to Bock and Moustaki, the value of the $a_k$ parameter may be used to rank the incorrect responses, with a higher value indicating a response that is more closely correlated with the latent trait and, therefore, better than a response with a lower value \cite{Bock2007}; however, the meaning of the $a_k$ and $b_k$ parameters is not as easily interpreted as the $a$ and $b$ of the 2PL model \cite{Thissen2010}. One shortcoming of the NRM is that the parameters are not uniquely defined, and a normalization constraint is required. This is often accomplished by setting the value of both parameters associated with one particular response to be fixed (at 0 or 1) and determining all other parameters relative to those. The NRM is excellent for analyzing data for which no prior information is available regarding the relative correctness of any of the responses; however, we have found that for our FMCE data, the parameters occasionally become reversed with choosing the correct response being associated with having a low value of $\theta$ (i.e., a poor overall understanding of Newtonian mechanics).

\section{Ranking Incorrect Responses}
\label{sec:irtrank}
In order to rank incorrect responses while properly accounting for the correct response, we use the 2PL-NRM nested logit model developed by Suh and Bolt \cite{Suh2010}. In this model, the probability of a student choosing a specific incorrect response $k$ is given by
\begin{align} 
	P_k(\theta)&=\left(1-\frac{1}{1+e^{-a\left(\theta-b\right)}}\right)\frac{e^{a_k\left(\theta-b_k\right)}}{\sum\limits_{i\neq\textrm{correct}} e^{a_i\left(\theta-b_i\right)}}
	\label{2plnrm}
\end{align}
where the parenthetical term is the probability of being not correct from the 2PL model, and the second term is Bock's NRM with the summation being over only the \textit{in}correct responses. In this model, the values of all $\theta$, $a$, and $b$ parameters are calculated using the 2PL model, and all $a_k$ and $b_k$ parameters are determined using the NRM, given the 2PL results. We used the A Multidimensional Item Response Theory (mirt) package in the R programming language to perform all IRT analyses \cite{r,Chalmers2012,mirt}.

To determine the ranking of incorrect responses, we calculated the values of $a$ and $b$ for every question, and $a_k$ and $b_k$ for each incorrect response choice. According to de Ayala, a data set must have at least 10 times as many response sets as the number of parameters to be calculated for an IRT model to have good convergence \cite{deAyala2008}; with our data set of $N=12,388$ response sets we are more than able to determine the 722 necessary parameters. 

We choose to omit response J from our ranking of incorrect responses because it does not correspond to a unique choice that students make about what they think is correct. Two students who choose response A (for example) agree that the information associated with A is correct, but two students who choose response J may or may not agree with each other regarding what a correct response would be. As such, we do not think it is valid to suggest that choosing J represents a unique level of correctness.

As a result of using the mirt package to apply the 2PL-NRM nested logit model, every response choice within each question has a unique $a_k$ value, implying that all answers are meaningfully different from each other. The question is then whether or not any of the $a_k$ values may be considered approximately equal to others, indicating approximately equal correlations with the $\theta$ parameter (i.e., response choices that are equally correct). To determine whether or not response choices are different from each other, we calculated the sampling distribution of values for each $a_k$ parameter by selecting random sample of 7,300 respondents using the sample function in R, and we used the mirt package to calculate each parameter \footnote{We chose 7,300 response sets to ensure we had at least 10 times as many response sets as parameters, as recommended in Ref.\ \cite{deAyala2008}. Samples were created without replacement.}. We repeated this process over 100,000 times to create a set of values for each parameter. 

The mirt package uniquely determines each value of $a_k$ by setting one parameter equal to 1 for each question \cite{mirt}. In order to ensure that we obtained a distribution of values for all parameters of interest, we chose to include one set of responses that included a ``dummy'' response and set $a_{0} = 0$ for this response. All other $a_k$ parameters are determined relative to $a_{0}$; as such, the $a_k$ values are only meaningful when compared within the same question, and there are no thresholds for determining whether a particular value of $a_k$ is high or low in and of itself. 

Using the effsize package, we calculated a Hedges' $g$ effect size to quantify the magnitude of the difference between the $a_k$ values for each pair \cite{effsize}. In our full data set, every response to every question is selected in at least one response set. The $a_k$ values reported in Tables \ref{rankTable126} and \ref{rankTable2747} are those determined from the full data set, with the dummy response set included to uniquely determine each parameter. Given that the values of $a_k$ parameters are only meaningful in relation to other values for the same question, the inclusion of the dummy response set has minimal impact on the overall results.

\begin{figure}
    \centering
    \includegraphics[width = \columnwidth]{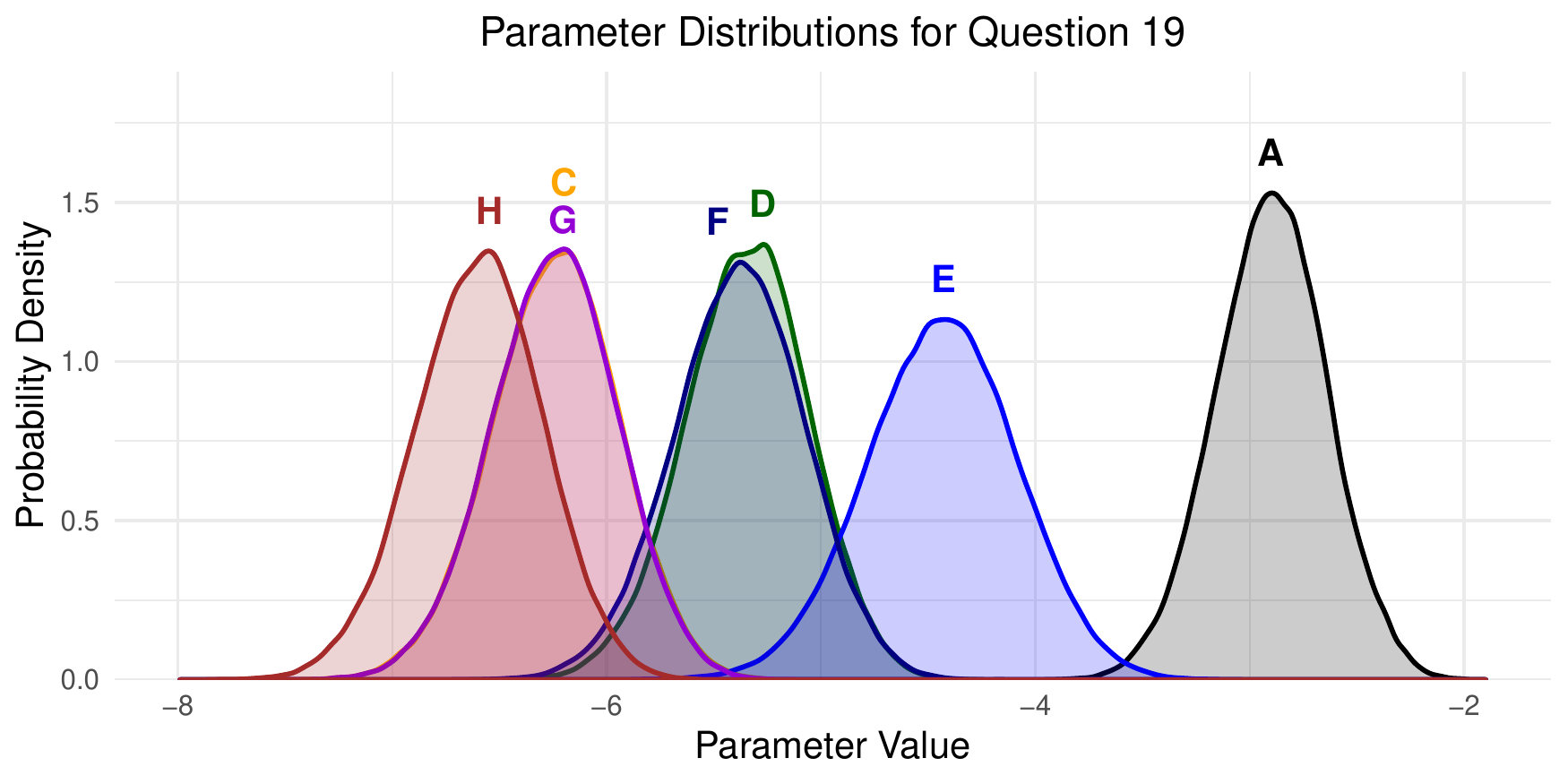}
    \caption{Distributions of $a_k$ parameters for question 19. Plot created using the geom\_density function in the ggplot2 package, and the directlabels package for R \cite{ggplot2,directlabels}.}
    \label{Q19ak}
\end{figure}

Figure \ref{Q19ak} shows a graphical representation of the parameter distributions for question 19. There are several key features to notice about these distributions:
\begin{itemize}
    \item there is no distribution for the correct response because the 2PL-NRM does not calculate an $a_k$ parameter for the correct response; it is automatically assumed to be the best,
    \item the distribution for $a_A$ is higher than any other value of $a_k$, with only minimal overlap with $a_E$, indicating that A has the highest parameter, and is thus the best incorrect response, and
    \item the distributions $a_C$ and $a_G$ are practically identical, indicating that these parameters have very similar values; thus, we would interpret them as being equally correct.
\end{itemize}
Other comparisons between various responses are a bit more ambiguous. The $a_D$ and $a_F$ distributions look quite similar, but not as similar as $a_C$ and $a_G$. The $a_H$ distribution is noticeably shifted to the left of $a_C$ and $a_G$, but there is still quite a bit of overlap. We use Hedges' $g$ to quantify the magnitude of the difference between each pair of distributions of the $a_k$ values: if $g$ is small ($g<0.5$) we conclude that the parameter values are effectively equal and the responses are equally correct, if $g$ is very large ($g\geq 1.3$) we conclude that the parameters are significantly different and that the responses represent different levels of understanding, and if $0.5\leq g < 1.3$ then we cannot make a conclusive determination.

\begin{table*}[h!]
\caption{Ranking results from IRT analyses (questions 1--26). These results show the complete ranking for every response choice to every question, as well as the associated $a_k$ values for each response, the value of Hedges' $g$ used to determine the ranking between each pair of parameters, and the percent of the data set that chose each response (rounded to the nearest percent).}
\begin{ruledtabular}
{\scriptsize \renewcommand{\arraystretch}{0.62} 
\begin{tabular}{rcc*{12}{c@{\hspace{-0.1em}}}c@{\hspace{7mm}}rcc*{12}{c@{\hspace{-0.1em}}}c}
Q1: &B&$>$&D&$>$&A&$\geq$&C&$=$&E&$=$&F&$\geq$&G&&&Q14: &E&$>$&H&$>$&F&$>$&G&$>$&A&$\geq$&D&$=$&C&$=$&B\\
$a_k$:&&&0.77&$>$&-0.18&$\geq$&-0.48&$\approx$&-0.62&$\approx$&-0.77&$\geq$&-1.39&&&$a_k$:&&&2.07&$>$&1.01&$>$&0.3&$>$&-0.55&$\geq$&-0.84&$\approx$&-0.97&$\approx$&-0.99\\
$g$:&&&&2.75&&1.25&&0.34&&0.28&&1.21&&&&$g$:&&&&2.7&&1.34&&1.89&&0.86&&0.32&&0.05&\\
\%:&32&&0&&66&&1&&0&&0&&0&&&\%:&27&&2&&0&&0&&62&&1&&5&&1\\[2ex]
Q2: &D&$>$&C&$>$&E&$\geq$&G&$\geq$&A&$=$&B&$>$&F&&&Q15: &E&$>$&G&$\geq$&D&$\geq$&H&$=$&C&$=$&F&$>$&A&$\geq$&B\\
$a_k$:&&&1.62&$>$&0.54&$\geq$&-0.14&$\geq$&-0.46&$\approx$&-0.51&$>$&-1.46&&&$a_k$:&&&0.94&$\geq$&0.43&$\geq$&0.19&$\approx$&0.17&$\approx$&0.04&$>$&-0.35&$\geq$&-0.62\\
$g$:&&&&2.69&&1.04&&0.54&&0.16&&2.86&&&&$g$:&&&&1.27&&0.69&&0.03&&0.44&&1.37&&1.07&\\
\%:&29&&4&&0&&0&&1&&65&&0&&&\%:&95&&0&&0&&0&&0&&0&&1&&2\\[2ex]
Q3: &F&$>$&B&$>$&E&$\geq$&G&$>$&A&$=$&C&$=$&D&&&Q16: &A&$>$&E&$>$&B&$>$&F&$=$&C&$>$&D&$=$&G&$\geq$&H\\
$a_k$:&&&0.59&$>$&0.29&$\geq$&0.13&$>$&-0.49&$\approx$&-0.61&$\approx$&-0.71&&&$a_k$:&&&0.94&$>$&0.53&$>$&-0.26&$\approx$&-0.36&$>$&-0.95&$\approx$&-0.96&$\geq$&-1.3\\
$g$:&&&&1.32&&0.73&&2.71&&0.45&&0.47&&&&$g$:&&&&1.5&&2.49&&0.34&&2.83&&0.05&&0.99&\\
\%:&41&&1&&6&&7&&1&&41&&3&&&\%:&31&&0&&1&&0&&66&&1&&0&&0\\[2ex]
Q4: &F&$>$&D&$>$&B&$=$&E&$\geq$&C&$\geq$&G&$>$&A&&&Q17: &E&$>$&F&$=$&B&$>$&D&$\geq$&G&$=$&C&$\geq$&H&$\geq$&A\\
$a_k$:&&&0.76&$>$&0.27&$\approx$&0.15&$\geq$&-0.02&$\geq$&-0.3&$>$&-0.7&&&$a_k$:&&&0.06&$\approx$&-0.02&$>$&-0.48&$\geq$&-0.64&$\approx$&-0.68&$\geq$&-0.81&$\geq$&-1.07\\
$g$:&&&&2.05&&0.5&&0.78&&1.28&&1.66&&&&$g$:&&&&0.25&&2.36&&0.76&&0.22&&0.58&&1.23&\\
\%:&31&&1&&1&&3&&0&&63&&1&&&\%:&24&&0&&53&&9&&2&&2&&5&&4\\[2ex]
Q5: &D&$>$&C&$>$&E&$>$&A&$\geq$&B&$=$&G&$=$&F&&&Q18: &B&$>$&A&$\geq$&E&$>$&D&$=$&G&$=$&H&$>$&C&$>$&F\\
$a_k$:&&&1.31&$>$&0.18&$>$&-0.45&$\geq$&-0.61&$\approx$&-0.62&$\approx$&-0.73&&&$a_k$:&&&1.07&$\geq$&0.77&$>$&-0.26&$\approx$&-0.35&$\approx$&-0.35&$>$&-0.67&$>$&-0.97\\
$g$:&&&&4.84&&2.56&&0.67&&0.06&&0.45&&&&$g$:&&&&1.26&&4.32&&0.39&&0&&1.46&&1.31&\\
\%:&48&&4&&1&&3&&36&&1&&7&&&\%:&27&&2&&1&&8&&9&&49&&1&&2\\[2ex]
Q6: &B&$>$&F&$>$&A&$=$&E&$\geq$&G&$>$&C&$>$&D&&&Q19: &B&$>$&A&$>$&E&$>$&D&$=$&F&$>$&C&$=$&G&$\geq$&H\\
$a_k$:&&&0.91&$>$&0.17&$\approx$&0.16&$\geq$&0.05&$>$&-0.36&$>$&-0.87&&&$a_k$:&&&2.31&$>$&0.76&$>$&-0.16&$\approx$&-0.19&$>$&-1.05&$\approx$&-1.06&$\geq$&-1.41\\
$g$:&&&&3.62&&0.06&&0.53&&1.98&&2.46&&&&$g$:&&&&4.98&&2.83&&0.12&&2.85&&0.01&&1.19&\\
\%:&19&&12&&7&&16&&6&&34&&5&&&\%:&27&&3&&0&&52&&1&&6&&5&&3\\[2ex]
Q7: &B&$>$&F&$=$&C&$=$&A&$>$&G&$>$&D&$=$&E&&&Q20: &G&$>$&B&$>$&A&$=$&D&$=$&C&$=$&E&$=$&H&$>$&F\\
$a_k$:&&&0.22&$\approx$&0.14&$\approx$&0.1&$>$&-0.39&$>$&-0.68&$\approx$&-0.78&&&$a_k$:&&&0.59&$>$&0.27&$\approx$&0.22&$\approx$&0.21&$\approx$&0.21&$\approx$&0.19&$>$&-0.23\\
$g$:&&&&0.34&&0.18&&2.23&&1.33&&0.43&&&&$g$:&&&&1.57&&0.25&&0.04&&0.03&&0.1&&2.31&\\
\%:&42&&4&&6&&8&&2&&4&&32&&&\%:&29&&0&&1&&1&&0&&1&&2&&64\\[2ex]
Q8: &A&$>$&D&$>$&B&$=$&E&$=$&C&$>$&G&$>$&F&&&Q21: &E&$>$&B&$>$&A&$>$&D&$=$&H&$=$&F&$\geq$&C&$>$&G\\
$a_k$:&&&1.1&$>$&0.48&$\approx$&0.37&$\approx$&0.35&$>$&-0.17&$>$&-0.94&&&$a_k$:&&&0.56&$>$&-0.01&$>$&-0.38&$\approx$&-0.47&$\approx$&-0.5&$\geq$&-0.69&$>$&-1\\
$g$:&&&&2.5&&0.43&&0.08&&2.08&&3.14&&&&$g$:&&&&2.86&&1.77&&0.43&&0.19&&0.87&&1.52&\\
\%:&23&&1&&4&&9&&2&&44&&16&&&\%:&36&&2&&8&&2&&25&&15&&2&&7\\[2ex]
Q9: &A&$>$&C&$=$&B&$>$&D&$\geq$&G&$=$&E&$=$&F&&&Q22: &A&$>$&C&$\geq$&B&$>$&D&$>$&G&$>$&F&$=$&E&&\\
$a_k$:&&&0.29&$\approx$&0.27&$>$&-0.17&$\geq$&-0.5&$\approx$&-0.55&$\approx$&-0.63&&&$a_k$:&&&0.76&$\geq$&0.57&$>$&0.17&$>$&-0.4&$>$&-1.1&$\approx$&-1.16&&\\
$g$:&&&&0.07&&1.85&&1.27&&0.16&&0.34&&&&$g$:&&&&0.91&&1.6&&1.93&&2.12&&0.09&&&\\
\%:&26&&1&&3&&67&&1&&3&&1&&&\%:&55&&1&&1&&0&&0&&0&&42&&\\[2ex]
Q10: &A&$>$&E&$\geq$&D&$>$&G&$\geq$&B&$=$&C&$=$&F&&&Q23: &B&$>$&A&$=$&C&$>$&F&$=$&G&$=$&E&$\geq$&D&&\\
$a_k$:&&&0.76&$\geq$&0.45&$>$&-0.2&$\geq$&-0.37&$\approx$&-0.4&$\approx$&-0.42&&&$a_k$:&&&0.59&$\approx$&0.58&$>$&-0.61&$\approx$&-0.69&$\approx$&-0.74&$\geq$&-0.99&&\\
$g$:&&&&1.2&&2.53&&0.65&&0.11&&0.08&&&&$g$:&&&&0.05&&5.67&&0.38&&0.27&&1.03&&&\\
\%:&35&&1&&2&&1&&54&&6&&1&&&\%:&43&&2&&1&&8&&41&&1&&4&&\\[2ex]
Q11: &A&$>$&D&$>$&B&$\geq$&C&$>$&E&$>$&G&$>$&F&&&Q24: &C&$>$&D&$>$&E&$=$&F&$>$&B&$>$&G&$>$&A&&\\
$a_k$:&&&0.53&$>$&0.22&$\geq$&0.11&$>$&-0.19&$>$&-0.46&$>$&-1.64&&&$a_k$:&&&-0.2&$>$&-0.67&$\approx$&-0.68&$>$&-0.91&$>$&-1.17&$>$&-1.61&&\\
$g$:&&&&1.5&&0.58&&1.56&&1.38&&6.23&&&&$g$:&&&&2.22&&0&&1.41&&1.51&&2.34&&&\\
\%:&34&&1&&4&&3&&4&&48&&7&&&\%:&53&&1&&1&&4&&33&&4&&3&&\\[2ex]
Q12: &A&$>$&B&$=$&C&$>$&F&$>$&D&$>$&E&$\geq$&G&&&Q25: &B&$>$&A&$>$&C&$>$&D&$=$&F&$>$&E&$\geq$&G&&\\
$a_k$:&&&0.23&$\approx$&0.16&$>$&-0.18&$>$&-0.48&$>$&-0.81&$\geq$&-1.14&&&$a_k$:&&&1.08&$>$&0.5&$>$&-0.65&$\approx$&-0.69&$>$&-1.14&$\geq$&-1.45&&\\
$g$:&&&&0.32&&1.37&&1.41&&1.6&&1.15&&&&$g$:&&&&2.47&&4.37&&0.13&&1.97&&1.24&&&\\
\%:&36&&2&&1&&0&&60&&1&&0&&&\%:&40&&6&&1&&1&&41&&5&&4&&\\[2ex]
Q13: &A&$>$&D&$=$&E&$>$&C&$\geq$&B&$\geq$&G&$\geq$&F&&&Q26: &C&$>$&G&$\geq$&D&$>$&F&$\geq$&B&$>$&E&$>$&A&&\\
$a_k$:&&&0.45&$\approx$&0.42&$>$&-0.63&$\geq$&-0.75&$\geq$&-0.87&$\geq$&-1.06&&&$a_k$:&&&0.54&$\geq$&0.42&$>$&-0.12&$\geq$&-0.35&$>$&-0.78&$>$&-1.32&&\\
$g$:&&&&0.11&&4.8&&0.57&&0.57&&0.84&&&&$g$:&&&&0.55&&2.31&&0.9&&1.79&&2.37&&&\\
\%:&44&&1&&1&&3&&50&&1&&1&&&\%:&55&&1&&1&&0&&2&&3&&37&&\\
\end{tabular}
}
\end{ruledtabular}
\label{rankTable126}
\end{table*}

\begin{table*}[h!]
\caption{Ranking results from IRT analyses continued (questions 27--47).}
\begin{ruledtabular}
{\scriptsize \renewcommand{\arraystretch}{0.67} 
\begin{tabular}{rcc*{12}{c@{\hspace{-0.1em}}}c@{\hspace{7mm}}rcc*{12}{c@{\hspace{-0.1em}}}c}
Q27: &A&$>$&D&$\geq$&B&$\geq$&E&$\geq$&C&$>$&G&$>$&F&&&Q38: &A&$>$&E&$=$&B&$\geq$&C&$\geq$&D&&&&&&\\
$a_k$:&&&0.3&$\geq$&0.06&$\geq$&-0.11&$\geq$&-0.31&$>$&-0.92&$>$&-2.1&&&$a_k$:&&&0.27&$\approx$&0.16&$\geq$&0.05&$\geq$&-0.2&&&&&&\\
$g$:&&&&1.17&&0.92&&1.06&&3.3&&6.13&&&&$g$:&&&&0.47&&0.53&&1.09&&&&&&&\\
\%:&42&&1&&5&&4&&7&&36&&5&&&\%:&23&&2&&66&&5&&3&&&&&&\\[2ex]
Q28: &A&$>$&E&$\geq$&B&$=$&C&$>$&F&$>$&D&$>$&G&&&Q39: &E&$>$&A&$\geq$&C&$=$&D&$>$&B&&&&&&\\
$a_k$:&&&0.31&$\geq$&0.16&$\approx$&0.15&$>$&-0.23&$>$&-0.61&$>$&-0.98&&&$a_k$:&&&-0.12&$\geq$&-0.4&$\approx$&-0.42&$>$&-1.28&&&&&&\\
$g$:&&&&0.64&&0.07&&1.55&&1.52&&1.4&&&&$g$:&&&&1.25&&0.07&&3.96&&&&&&&\\
\%:&38&&2&&2&&1&&0&&57&&0&&&\%:&51&&3&&4&&35&&6&&&&&&\\[2ex]
Q29: &A&$>$&E&$>$&D&$>$&C&$>$&F&$=$&G&$=$&B&&&Q40: &A&$>$&E&$\geq$&B&$=$&F&$\geq$&C&$\geq$&G&$\geq$&H&$\geq$&D\\
$a_k$:&&&1.07&$>$&0.17&$>$&-0.46&$>$&-0.81&$\approx$&-0.85&$\approx$&-0.9&&&$a_k$:&&&0.29&$\geq$&0.12&$\approx$&0.08&$\geq$&-0.18&$\geq$&-0.4&$\geq$&-0.6&$\geq$&-0.84\\
$g$:&&&&3.62&&2.34&&1.37&&0.16&&0.16&&&&$g$:&&&&0.77&&0.22&&1.09&&0.84&&0.55&&0.72&\\
\%:&45&&5&&1&&3&&6&&1&&40&&&\%:&88&&2&&2&&0&&1&&0&&0&&7\\[2ex]
Q30: &E&$>$&B&$\geq$&F&$\geq$&C&$=$&A&$\geq$&D&&&&&Q41: &F&$>$&E&$>$&B&$>$&D&$=$&A&$\geq$&C&$>$&G&$\geq$&H\\
$a_k$:&&&0.25&$\geq$&0.1&$\geq$&-0.1&$\approx$&-0.22&$\geq$&-0.38&&&&&$a_k$:&&&0.46&$>$&0.06&$>$&-0.28&$\approx$&-0.36&$\geq$&-0.48&$>$&-0.87&$\geq$&-0.99\\
$g$:&&&&0.58&&0.74&&0.44&&0.62&&&&&&$g$:&&&&2.05&&1.69&&0.39&&0.55&&1.98&&0.56&\\
\%:&42&&1&&1&&1&&54&&1&&&&&\%:&72&&1&&7&&1&&1&&5&&7&&5\\[2ex]
Q31: &E&$>$&D&$=$&F&$=$&C&$\geq$&A&$\geq$&B&&&&&Q42: &B&$>$&E&$>$&G&$=$&F&$=$&D&$=$&H&$\geq$&A&$=$&C\\
$a_k$:&&&-0.01&$\approx$&-0.04&$\approx$&-0.14&$\geq$&-0.39&$\geq$&-0.61&&&&&$a_k$:&&&0.06&$>$&-0.37&$\approx$&-0.38&$\approx$&-0.43&$\approx$&-0.44&$\geq$&-0.61&$\approx$&-0.69\\
$g$:&&&&0.09&&0.41&&0.95&&0.85&&&&&&$g$:&&&&1.9&&0.02&&0.26&&0.06&&0.79&&0.42&\\
\%:&47&&1&&31&&1&&5&&14&&&&&\%:&78&&1&&0&&1&&2&&7&&2&&7\\[2ex]
Q32: &E&$>$&F&$=$&A&$=$&D&$=$&B&$\geq$&C&&&&&Q43: &D&$>$&B&$\geq$&G&$=$&C&$=$&F&$=$&A&$=$&H&$>$&E\\
$a_k$:&&&-0.06&$\approx$&-0.16&$\approx$&-0.27&$\approx$&-0.37&$\geq$&-0.5&&&&&$a_k$:&&&0.03&$\geq$&-0.09&$\approx$&-0.12&$\approx$&-0.21&$\approx$&-0.26&$\approx$&-0.27&$>$&-0.77\\
$g$:&&&&0.41&&0.44&&0.39&&0.55&&&&&&$g$:&&&&0.54&&0.14&&0.41&&0.19&&0.06&&2.1&\\
\%:&44&&7&&5&&2&&39&&2&&&&&\%:&89&&1&&0&&1&&0&&4&&1&&1\\[2ex]
Q33: &E&$>$&D&$\geq$&F&$\geq$&B&$\geq$&C&$\geq$&A&&&&&Q44: &B&$>$&D&$=$&C&$\geq$&A&&&&&&&&\\
$a_k$:&&&0.2&$\geq$&0.08&$\geq$&-0.14&$\geq$&-0.29&$\geq$&-0.45&&&&&$a_k$:&&&-0.03&$\approx$&-0.04&$\geq$&-0.25&&&&&&&&\\
$g$:&&&&0.56&&1.03&&0.73&&0.7&&&&&&$g$:&&&&0.06&&0.81&&&&&&&&&\\
\%:&93&&1&&1&&1&&2&&2&&&&&\%:&44&&2&&4&&49&&&&&&&&\\[2ex]
Q34: &E&$>$&F&$=$&A&$=$&D&$=$&C&$=$&B&&&&&Q45: &B&$>$&C&$=$&A&$=$&D&&&&&&&&\\
$a_k$:&&&-0.13&$\approx$&-0.14&$\approx$&-0.23&$\approx$&-0.24&$\approx$&-0.35&&&&&$a_k$:&&&-0.17&$\approx$&-0.21&$\approx$&-0.24&&&&&&&&\\
$g$:&&&&0.01&&0.37&&0.06&&0.42&&&&&&$g$:&&&&0.14&&0.11&&&&&&&&&\\
\%:&41&&1&&3&&2&&1&&50&&&&&\%:&54&&6&&37&&3&&&&&&&&\\[2ex]
Q35: &A&$>$&E&$>$&B&$=$&C&$\geq$&D&&&&&&&Q46: &A&$>$&B&$=$&D&$\geq$&C&&&&&&&&\\
$a_k$:&&&0.77&$>$&-0.13&$\approx$&-0.25&$\geq$&-0.57&&&&&&&$a_k$:&&&0.04&$\approx$&-0.02&$\geq$&-0.3&&&&&&&&\\
$g$:&&&&3.6&&0.47&&1.27&&&&&&&&$g$:&&&&0.24&&1.11&&&&&&&&&\\
\%:&56&&1&&38&&2&&2&&&&&&&\%:&44&&16&&17&&22&&&&&&&&\\[2ex]
Q36: &A&$>$&E&$>$&C&$\geq$&B&$>$&D&&&&&&&Q47: &A&$>$&D&$=$&B&$\geq$&C&&&&&&&&\\
$a_k$:&&&0.51&$>$&0.16&$\geq$&-0.01&$>$&-0.54&&&&&&&$a_k$:&&&0.01&$\approx$&-0.01&$\geq$&-0.26&&&&&&&&\\
$g$:&&&&1.51&&0.72&&2.32&&&&&&&&$g$:&&&&0.08&&1.03&&&&&&&&&\\
\%:&23&&1&&68&&3&&3&&&&&&&\%:&46&&13&&22&&19&&&&&&&&\\[2ex]
Q37: &A&$>$&E&$>$&B&$>$&C&$\geq$&D&&&&&&&\\
$a_k$:&&&0.29&$>$&-0.09&$>$&-0.56&$\geq$&-0.78&&&&&&&\\
$g$:&&&&1.84&&2.27&&1.07&&&&&&&&\\
\%:&65&&4&&3&&22&&4&&&&&&&\\
\end{tabular}
}
\end{ruledtabular}
% \hline\hline
\label{rankTable2747}
\end{table*}

Tables \ref{rankTable126} and \ref{rankTable2747} show the IRT ranking results for each question on the FMCE, including the $a_k$ value for each response, the value of Hedges' $g$ for each nearest-neighbor comparison, and the percentage of the data set that chose each response. Consider the ranking shown for question 19 as it relates to Fig.\ \ref{Q19ak}. The $a_C$ and $a_G$ parameters are nearly identical (when rounded to two decimal places), and the effect size between their distributions is negligibly small ($g=0.01$). The effect size between D and F is also quite small ($g=0.12$), suggesting that choosing either of these two responses indicates a similar level of understanding, and the effect size between G and H is large, but not above our threshold for different responses ($g=1.19$) \footnote{The effect size between C and H is also in the ambiguous range: $g=1.20$.}. All other effect sizes for question 19 are very large, indicating considerably different values of $a_k$ that correspond to different levels of understanding. 

Question 19 on the FMCE presents students with a situation in which a toy car ``moves toward the left and is speeding up at a steady rate (constant acceleration)'' and asks them to choose an appropriate graph of force vs.\ time. The correct response to question 19 is B: a graph with a constant negative value (zero slope). According to these results, the best incorrect answer is A: a graph with a constant positive value (zero slope). The second-best incorrect response is E: a graph with a constant zero value (zero slope). All other responses are graphs with nonzero slope. This suggests that realizing that a constant acceleration indicates a constant force is indicative of an above-average understanding of basic Newtonian mechanics. This result alone may not be revolutionary to anyone who has taught introductory mechanics, but the implication that claiming that zero force is required to make an object speed up is a better answer than selecting a graph showing a changing force may be more surprising. Response E is chosen by fewer than 1\% of the data set, but these students seem to otherwise have a fairly strong understanding of Newton's laws as measured by the FMCE.

\begin{table*}[bt]
\caption{IRT ranking results: filtered so that only responses given by at least 1\% of the population are included.}
\begin{ruledtabular}
{\scriptsize
\begin{tabular}{l*{15}{c}@{\hspace{5mm}}l*{15}{c}}
Q1: &B&$>$&A&&&&&&&&&&&&&Q25: &B&$>$&A&$>$&D&$=$&F&$>$&E&$\geq$&G&&&&\\[1ex]
Q2: &D&$>$&C&$>$&A&$=$&B&&&&&&&&&Q26: &C&$>$&G&$>$&B&$>$&E&$>$&A&&&&&&\\[1ex]
Q3: &F&$>$&E&$\geq$&G&$>$&C&$=$&D&&&&&&&Q27: &A&$>$&B&$\geq$&E&$\geq$&C&$>$&G&$>$&F&&&&\\[1ex]
Q4: &F&$>$&E&$>$&G&&&&&&&&&&&Q28: &A&$>$&E&$\geq$&B&$>$&D&&&&&&&&\\[1ex]
Q5: &D&$>$&C&$>$&E&$>$&A&$\geq$&B&$\geq$&F&&&&&Q29: &A&$>$&E&$>$&C&$>$&F&$=$&B&&&&&&\\[1ex]
Q6: &B&$>$&F&$>$&A&$=$&E&$\geq$&G&$>$&C&$>$&D&&&Q30: &E&$>$&B&$\geq$&F&$\geq$&C&$=$&A&$\geq$&D&&&&\\[1ex]
Q7: &B&$>$&F&$=$&C&$=$&A&$>$&G&$>$&D&$=$&E&&&Q31: &E&$>$&D&$=$&F&$=$&C&$\geq$&A&$\geq$&B&&&&\\[1ex]
Q8: &A&$>$&D&$>$&B&$=$&E&$=$&C&$>$&G&$>$&F&&&Q32: &E&$>$&F&$=$&A&$=$&D&$=$&B&$\geq$&C&&&&\\[1ex]
Q9: &A&$>$&B&$>$&D&$>$&E&&&&&&&&&Q33: &E&$>$&D&$\geq$&F&$>$&C&$\geq$&A&&&&&&\\[1ex]
Q10: &A&$>$&D&$>$&B&$=$&C&&&&&&&&&Q34: &E&$>$&F&$=$&A&$=$&D&$=$&C&$=$&B&&&&\\[1ex]
Q11: &A&$>$&B&$\geq$&C&$>$&E&$>$&G&$>$&F&&&&&Q35: &A&$>$&E&$>$&B&$=$&C&$\geq$&D&&&&&&\\[1ex]
Q12: &A&$>$&B&$>$&D&&&&&&&&&&&Q36: &A&$>$&E&$>$&C&$\geq$&B&$>$&D&&&&&&\\[1ex]
Q13: &A&$>$&C&$\geq$&B&$>$&F&&&&&&&&&Q37: &A&$>$&E&$>$&B&$>$&C&$\geq$&D&&&&&&\\[1ex]
Q14: &E&$>$&H&$>$&A&$>$&C&$=$&B&&&&&&&Q38: &A&$>$&E&$=$&B&$\geq$&C&$\geq$&D&&&&&&\\[1ex]
Q15: &E&$>$&A&$\geq$&B&&&&&&&&&&&Q39: &E&$>$&A&$\geq$&C&$=$&D&$>$&B&&&&&&\\[1ex]
Q16: &A&$>$&C&&&&&&&&&&&&&Q40: &A&$>$&E&$\geq$&B&$>$&C&$>$&D&&&&&&\\[1ex]
Q17: &E&$>$&B&$>$&D&$\geq$&G&$=$&C&$\geq$&H&$\geq$&A&&&Q41: &F&$>$&E&$>$&B&$>$&D&$\geq$&C&$>$&G&$\geq$&H&&\\[1ex]
Q18: &B&$>$&A&$>$&D&$=$&G&$=$&H&$>$&C&$>$&F&&&Q42: &B&$>$&E&$>$&D&$=$&H&$\geq$&A&$=$&C&&&&\\[1ex]
Q19: &B&$>$&A&$>$&D&$>$&C&$=$&G&$\geq$&H&&&&&Q43: &D&$>$&B&$\geq$&C&$\geq$&A&$=$&H&&&&&&\\[1ex]
Q20: &G&$>$&H&$>$&F&&&&&&&&&&&Q44: &B&$>$&D&$=$&C&$\geq$&A&&&&&&&&\\[1ex]
Q21: &E&$>$&B&$>$&A&$>$&D&$=$&H&$=$&F&$\geq$&C&$>$&G&Q45: &B&$>$&C&$=$&A&$=$&D&&&&&&&&\\[1ex]
Q22: &A&$>$&E&&&&&&&&&&&&&Q46: &A&$>$&B&$=$&D&$\geq$&C&&&&&&&&\\[1ex]
Q23: &B&$>$&A&$=$&C&$>$&F&$=$&G&$>$&D&&&&&Q47: &A&$>$&D&$=$&B&$\geq$&C&&&&&&&&\\[1ex]
Q24: &C&$>$&F&$>$&B&$>$&G&$>$&A&&&&&&&\\
\end{tabular}
}
\end{ruledtabular}
\label{rankTableFiltered}
\end{table*}

Table \ref{rankTableFiltered} shows IRT rankings that have been filtered to only include responses given by at least 1.00\% of the population. For some questions (such as 1, 16, and 22) the difference is quite stark, with only the correct and one incorrect answer choice remaining. For many of the questions (such as 2, 14, and 26) the rankings remain the same, but many of the responses that are seen as equivalent to others (or ambiguously ranked) have been eliminated. Moreover, a smaller fraction of the rankings in Table \ref{rankTableFiltered} are ``$\geq$'' as compared to Tables \ref{rankTable126} and \ref{rankTable2747} (29\% vs.\ 35\%), and a greater fraction of rankings are ``$>$'' (56\% vs.\ 36\%). This suggests that many of the ambiguities in rankings may be attributed to the relatively low probability of choosing those responses at all levels of physics understanding, which could result in relatively broad distributions of parameter values generated by randomly selecting subsets of data.

\section{Relating Response Rankings to IRT Plots}
\label{sec:rankplot}
\begin{figure*}[tb]
    \centering
    \begin{tabular}{ccc}
    \includegraphics[width = 0.3 \textwidth]{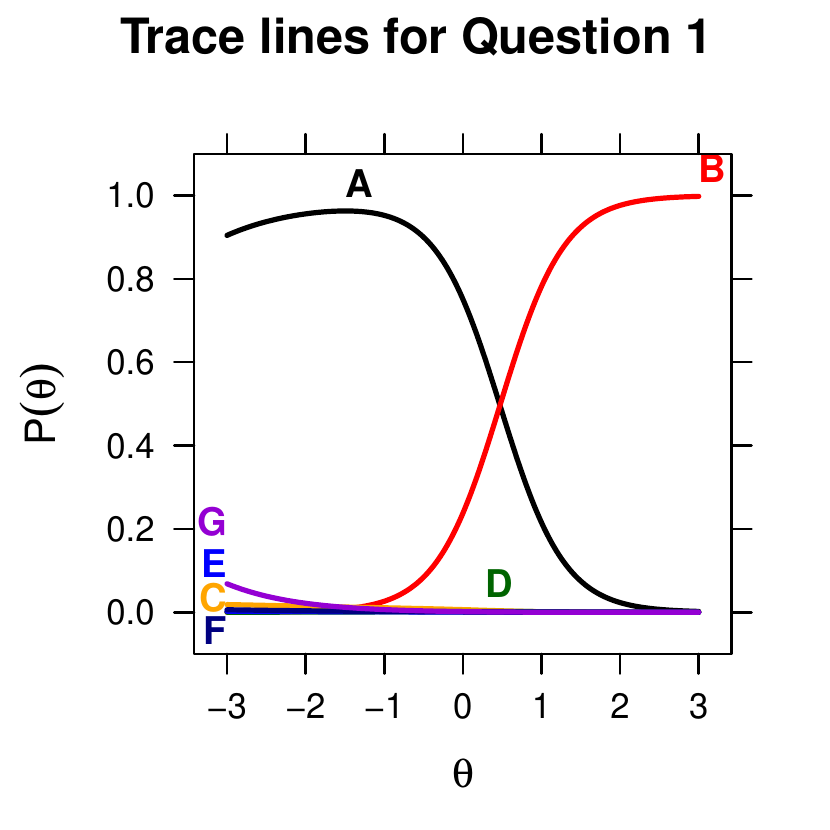}&
    \includegraphics[width = 0.3 \textwidth]{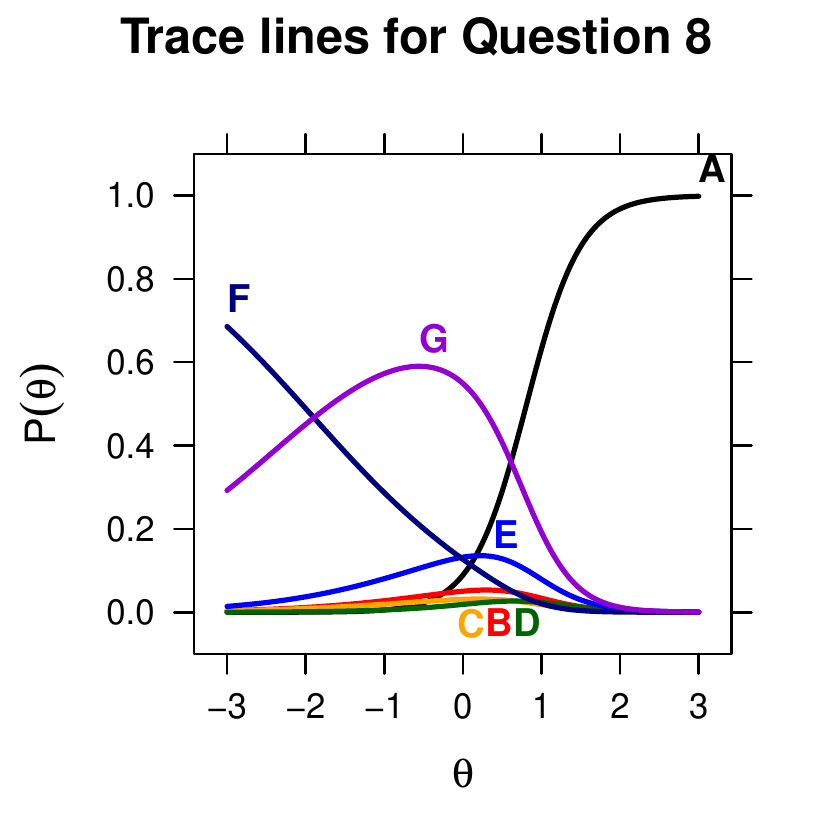}&
    \includegraphics[width = 0.3 \textwidth]{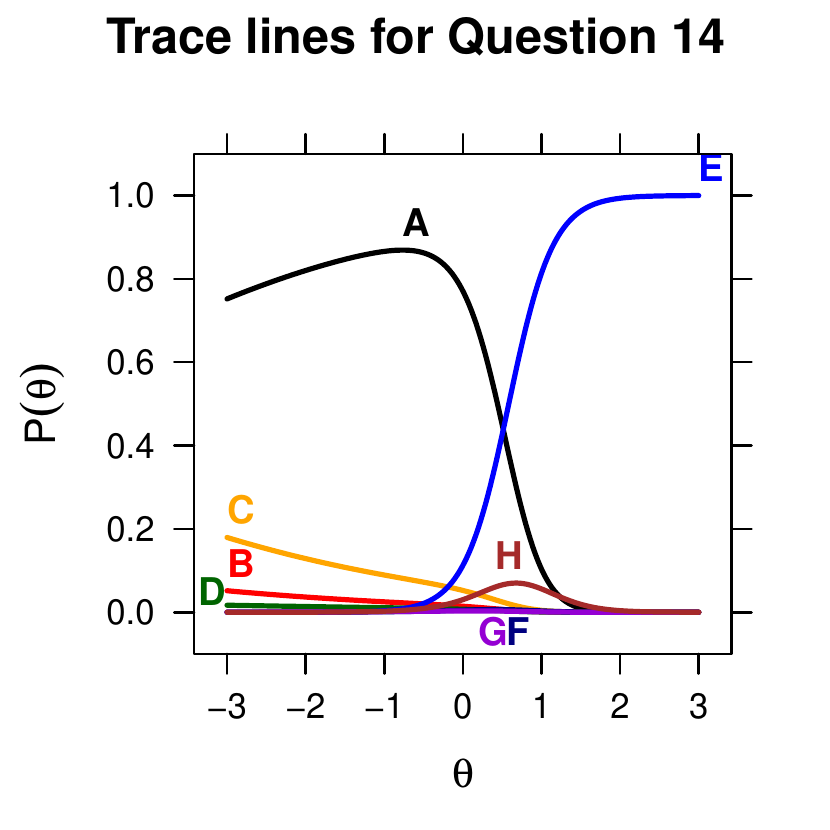}\\

    (a) Question 1 & (b) Question 8 & (c) Question 14 \\[5ex]

    \includegraphics[width = 0.3 \textwidth]{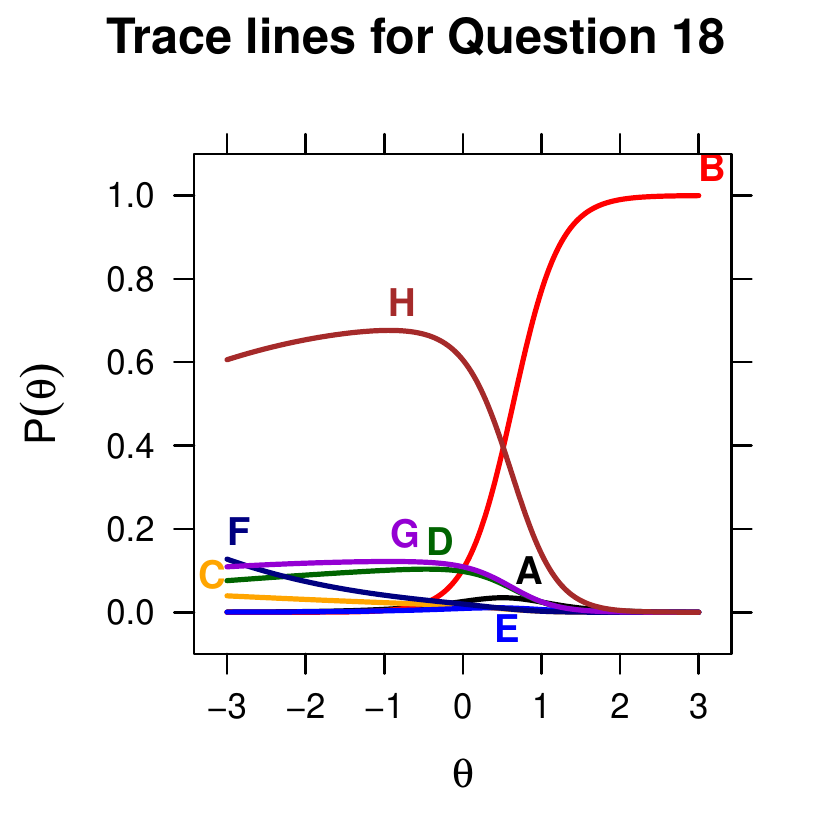}& 
    \includegraphics[width = 0.3 \textwidth]{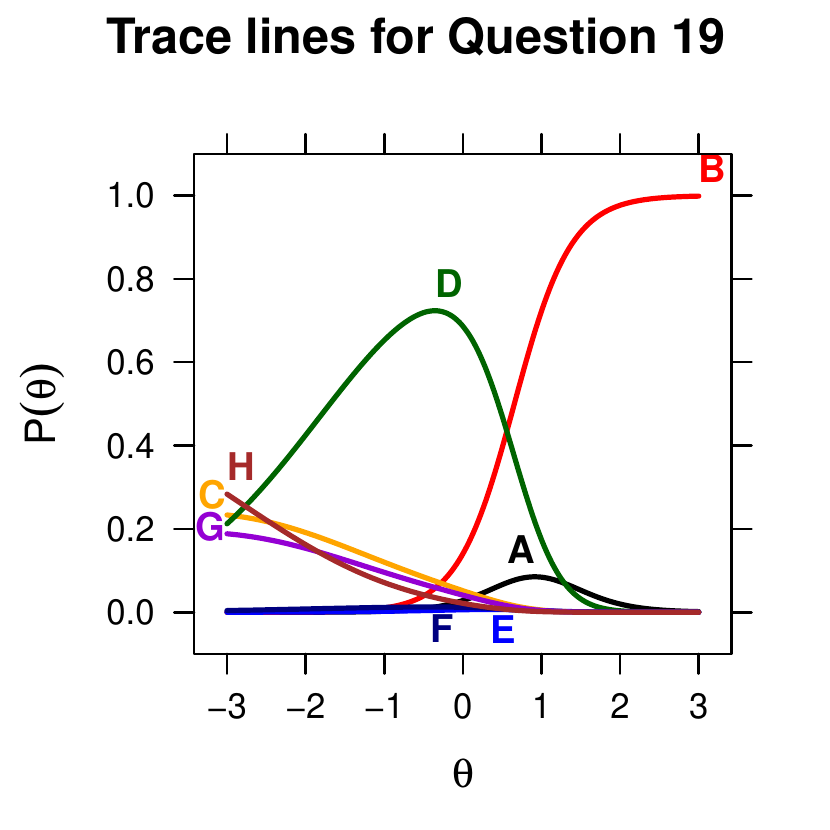}&
    \includegraphics[width = 0.3 \textwidth]{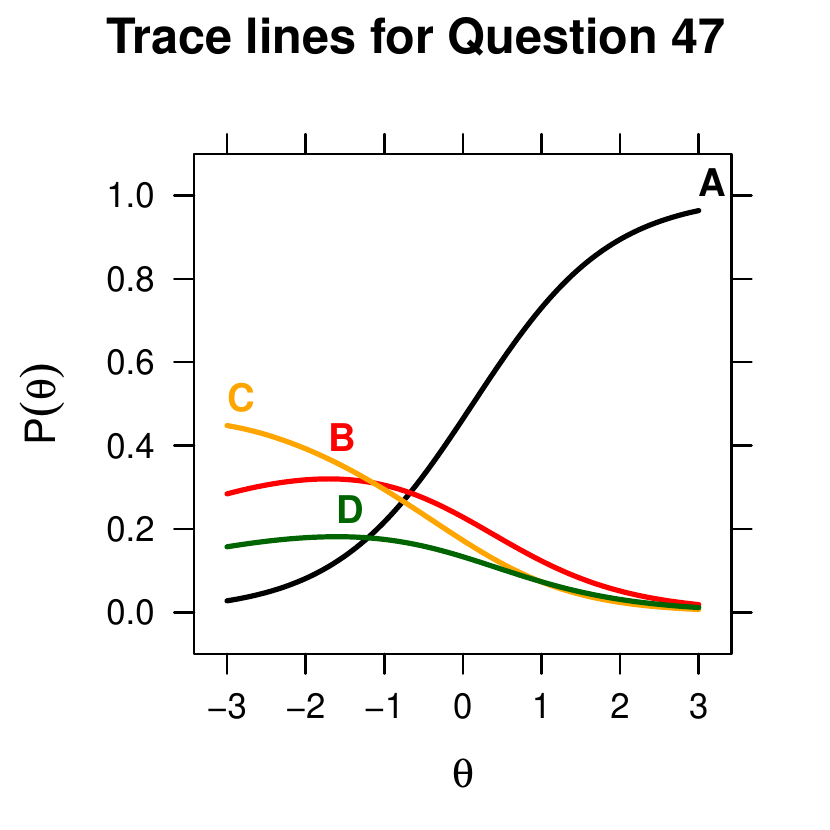}\\

    (d) Question 18 & (e) Question 19 & (f) Question 47\\
    \end{tabular}
    
    \caption{IRT plots from the 2PL-NRM nested logit model for questions 1, 8, 14, 18, 19, and 47: probability of choosing each response as a function of the person parameter $\theta$ (representing overall understanding of Newtonian mechanics).}
    \label{nrmPlots}
\end{figure*}

We can use plots of the IRT curves to better understand these rankings. Figure \ref{nrmPlots} shows the 2PL-NRM curves for every response to several question. (The 2PL-NRM IRT plots for all other questions are included in the appendix.) In the filtered ranking, Question 1 only includes two dominant responses. This is consistent with Fig.\ \ref{nrmPlots}(a) in which responses A and B dominate at all values of $\theta$ (overall understanding), and all other responses have near zero probability of being chosen. Figure \ref{nrmPlots}(c) shows that Q14 is a bit more interesting: there are still two dominant responses (the correct E and a single incorrect A), but less-common incorrect responses are chosen differently by students with different $\theta$ values. Students who choose response H are likely to have an above-average understanding ($\theta>0$) with the H curve having a roughly symmetric probability distribution centered around $\theta=b\approx0.75$, while response C is mostly chosen by students with below-average understanding and is more and more likely with lower values of $\theta$. This is consistent with the ranking in Table \ref{rankTableFiltered} with H being a better response than the most common A, and C being a worse response. We can also see in Fig.\ \ref{nrmPlots}(c) that responses B and C on Q14 have a similar shape, with the highest probability of choosing each being at the low end of the $\theta$-axis; this is consistent with these responses being considered equivalent in Table \ref{rankTableFiltered} even though the value of the probability is quite different for each. It is also important to notice that the curves for the correct responses (B for Q1 and E for Q14) are identical to the 2PL curves for the questions shown in Fig. \ref{2plPlots}.

    Question 19 in Fig.\ \ref{nrmPlots}(e) is very similar to Q14 in that there is a single most-common-incorrect response (D), a better response (A) that has a relatively narrow symmetric probability distribution centered around $\theta\approx b>0$, and some worse responses (C, G, H) that have their highest probabilities at the low end of the $\theta$-axis. The major differences between the plots for Q19 and Q14 are that the most-common-incorrect response has a much broader range of values in the $\theta<0$ regime for Q19. In fact, response D is approximately equally probable to responses C, G, and H around $\theta\approx-3$. Moreover, the probability curves for the equivalent responses C and G are basically on top of each other for the entire range of $\theta$. This is strong evidence that students choose these responses in roughly the same proportions, and it may indicate that students are choosing them for the same reasons. Response H also has similar probabilities to C and G, but has a distinctly more negative slope than either, indicating a greater likelihood of being chosen by students with lower $\theta$ values. 
    
    The plot for question 18 in Fig.\ \ref{nrmPlots}(d) is even more complex than that of Q14, but there are several similarities between them. Once again there is a single dominant incorrect response (H), but it is less dominant than the most common response to Q1 or Q14. This is largely due to G and D being equivalent to H (according to Tables \ref{rankTable126} and \ref{rankTableFiltered}), and all three sharing a similar shape in which the probability is relatively uniform for $\theta<0$ (but has a slight peak around $-1<\theta<-0.5$) and decreases to zero for $0<\theta< 2$. Responses D and G each have probabilities between about 0.1 and 0.15 for $\theta<0$, which accounts for the probability of the most common response H never rising above 0.75. Also in Fig.\ \ref{nrmPlots}(d) we can see a small bump in probability for the better-than-common response A around $\theta\approx0.75$, and the worst responses (C and F) are most common at the lowest values of $\theta$. Questions 14, 18, and 19 all indicate that students with above-average understanding ($\theta>0$) are more likely to choose both the correct response and the highest-ranked incorrect response than students with $\theta<0$.
    
    The 2PL-NRM plot for question 47 in Fig.\ \ref{nrmPlots}(f) shows an example of a question on the FMCE for which there is not a single dominant incorrect response. Once again we can see the ranking from Table \ref{rankTableFiltered} in the shape of the curves: A is correct, D and B have similar shapes with maximum probabilities around $\theta \approx -1.5$, and C with the highest probability at the low end of the $\theta$-axis with a distinctly negative slope. What makes Q47 really interesting is that none of the responses is uniformly zero over a broad range of $\theta$ values. All incorrect responses have probabilities above 0.15 for $-3<\theta<-1$, and the probabilities for two of the incorrect responses (B and C) are roughly equal to the correct response around $\theta\approx-0.5$ (with response D only about 0.10 lower). The relatively high probabilities of all incorrect answer choices on Q47 may contribute to the low value of the discrimination parameter $a$ (see Fig.\ \ref{2plPlots}).
    
    Question 8 (Fig.\ \ref{nrmPlots}(b)) shows an interesting example of a case where there are multiple responses (B, C, D, and E) ranked higher than the most common incorrect response G. Only response F is ranked lower than G (with $g\approx16$). On the FMCE, question 8 is the first in a set of three that asks students about the forces on a toy car as it rolls up and down a ramp. In question 8, the car is moving up and slowing down: response A is correct that the net force on the car is constant and down the ramp, response G is that the force is up the ramp and decreasing, while response F is that the force is up the ramp and increasing. Responses B and C both indicate a force down the ramp (increasing and decreasing, respectively), response D indicates zero net force, and response E is that the force is up the ramp and constant. All of these better-than-most-common responses agree with the correct answer in one way (either the direction of the force or the fact that it is constant), while G and F have nothing in common with the correct response. 
    
    From a visual perspective, the approximate value of $\theta$ for which the probability curve for a particular response is maximized indicates the ranking for that response \footnote{The label for each curve is placed as close to the peak as possible without overlapping with either another label or another curve.}. The curve for the correct answer is always a monotonically increasing function of $\theta$; answers that are better than a most-common incorrect answer (like H on Q14 in Fig.\ \ref{nrmPlots}(c) and A on Q18 in Fig.\ \ref{nrmPlots}(f)) have relatively narrow probability distributions with peaks around $\theta=b$; the worst response in each rank has its highest value at the low end of the the displayed $\theta$-axis with a distinctly negative slope, indicating that lower values of $\theta$ would yield even higher probabilities of choosing those responses. Plots of IRT probability curves for all FMCE questions are included in Figs.\ \ref{fig:2pl-nrm120}--\ref{fig:2pl-nrm4147} below.

\section{Anomalous Results}
\label{sec:anomalous}
\begin{figure}
    \centering
    \includegraphics[width = \columnwidth]{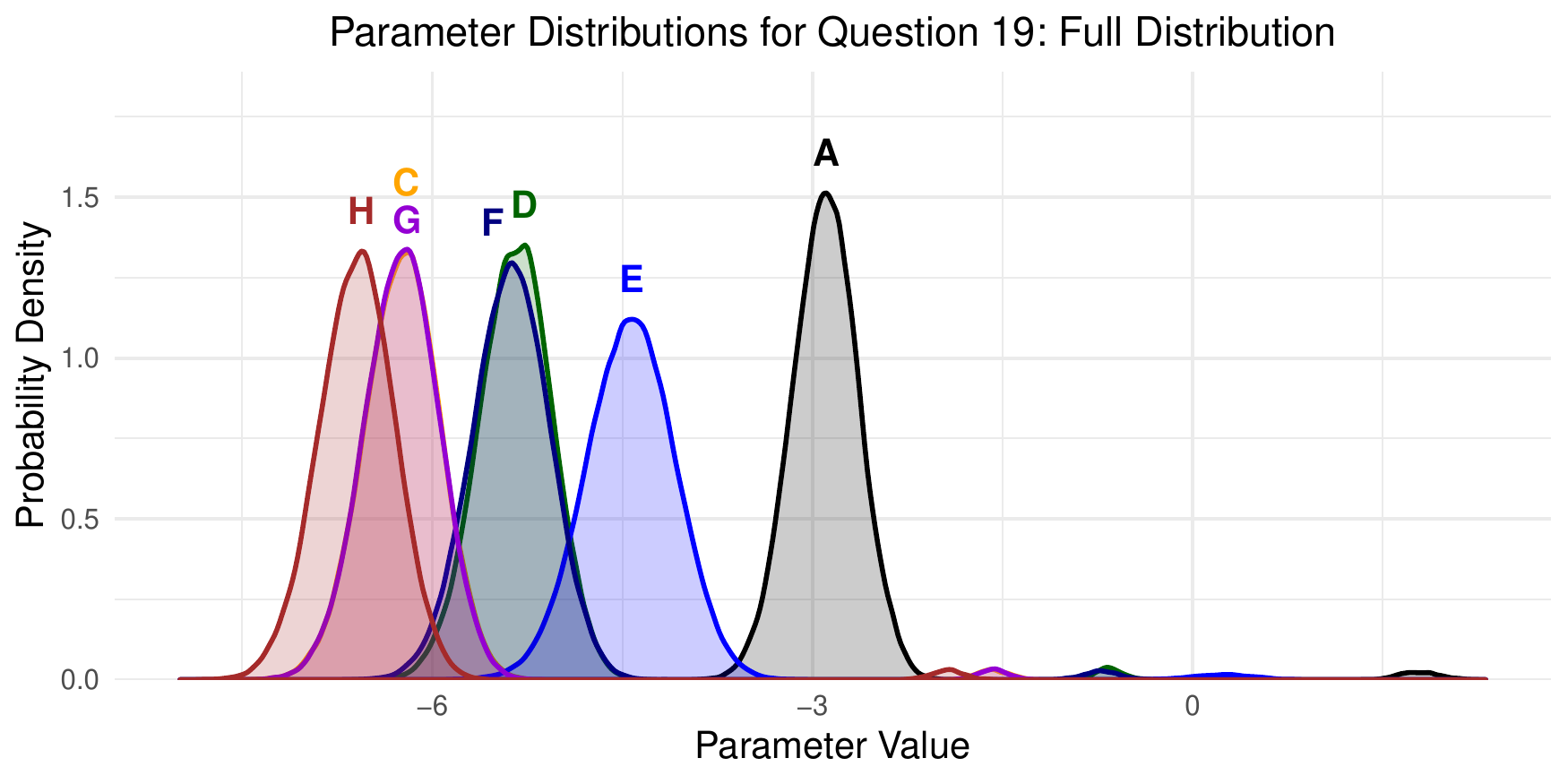}
    \caption{The complete parameter distribution for question 19. The large labeled curves on the left side of the plot contain 99\% of the results (also shown in Fig.\ \ref{Q19ak}), and the small bumps on the right side of the plot contain about 1\% of the results (also shown in Fig.\ \ref{Q19akOther}). A gap in the horizontal axis of about 1.5 exists between the left and right distributions for all parameters.}
    \label{Q19akAll}
\end{figure}

\begin{figure}
    \centering
    \includegraphics[width = \columnwidth]{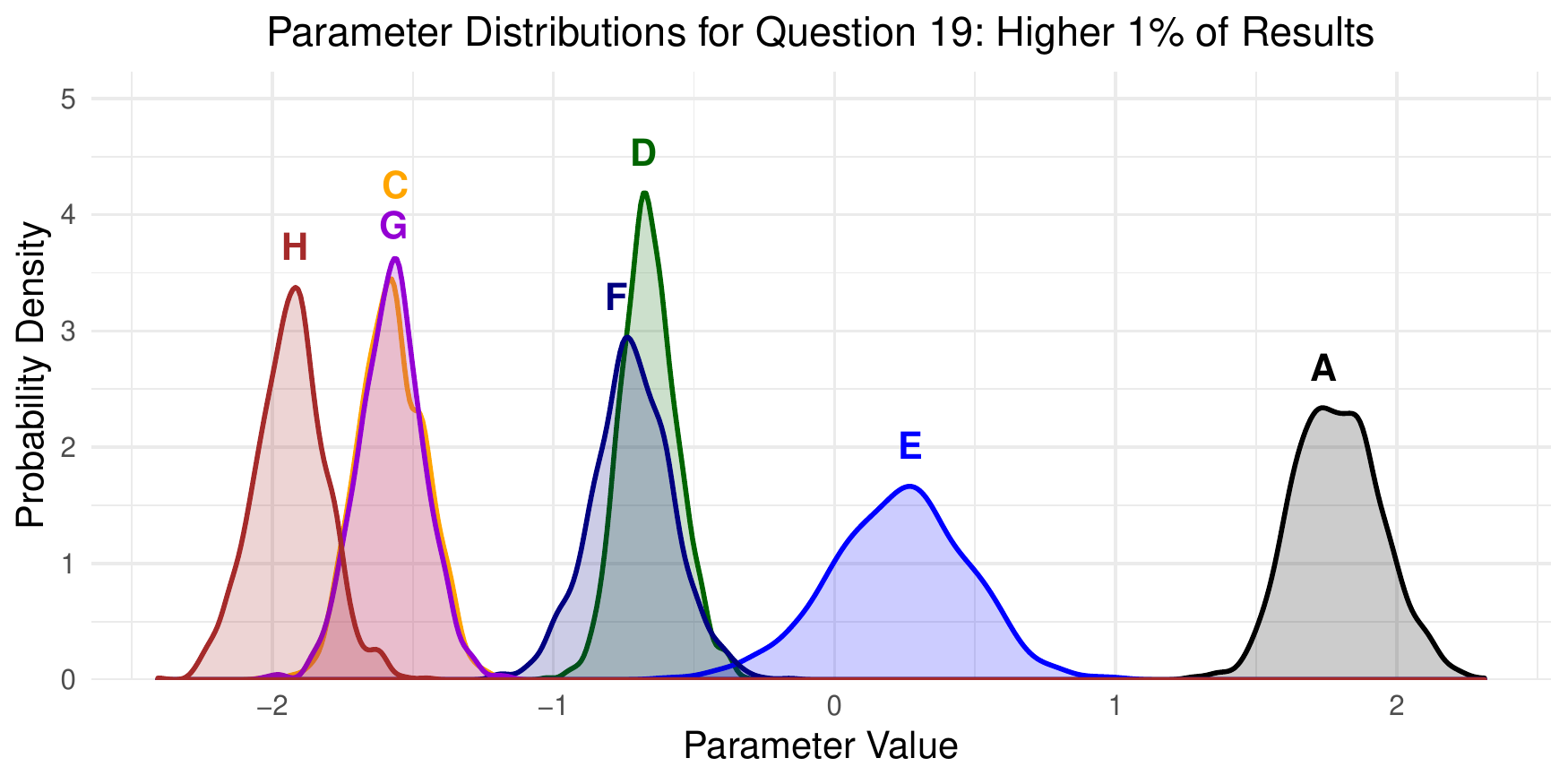}
    \caption{Enlarged version of the right side of Fig.\ \ref{Q19akAll}. These distributions represent about 1\% of the total results. The overall shapes and locations of the parameter distributions are similar to Fig.\ \ref{Q19ak}.}
    \label{Q19akOther}
\end{figure}

We used Rowan University's high-performance computing cluster to generate the distribution of values for each parameter. In examining the results, we noticed some anomalous values for each parameter as shown by the small bumps on the right side of Fig.\ \ref{Q19akAll}. These bumps represent about 1\% of the results. Figure \ref{Q19akOther} shows an enlarged version of the smaller distribution \footnote{The vertical axis in Figs.\ \ref{Q19ak}, \ref{Q19akAll}, and \ref{Q19akOther} is probability density, which is normalized based on the visible distribution; therefore, the values on the vertical axis should not be compared between plots.}.

Comparing Figs.\ \ref{Q19ak} and \ref{Q19akOther}, we see that the overall shapes and locations of the parameter distributions are similar. The distributions in Fig.\ \ref{Q19akOther} are not as smooth or as broad as those in Fig.\ \ref{Q19ak}, but this may be attributed to the sampling method. Possibly the most interesting feature of these distributions is that there is not overlap between the left and right distributions of any parameter (which is not clear in any of the figures). For example, consider the distribution of the parameter for response A in Fig.\ \ref{Q19akAll}: the left distribution for the parameter $a_A$ (also shown as the black curve in Fig.\ \ref{Q19ak}) contains values between $-4.2$ and $-1.9$, while the right distribution (also shown as the black curve in Fig.\ \ref{Q19akOther}) contains values between $+1.3$ and $+2.3$. This leaves a gap of $3.2$ that contains no results.

We found the same anomalous results in the distributions for all parameters in all questions. In 1,003 out of the 105,600 analyses all of the $a_k$ parameters are about 2--5 units larger than in the other 104,597 analyses. The same 1,003 analyses provide the higher results for all parameters, and for all questions the distributions of the higher values are similar in shape and relative location to the distributions for the lower values. Given the similarities in the distributions, and the consistency of the same 1\% of analyses providing the differences for all parameters, we feel comfortable using the results from the 104,597 main analyses to make claims about the differences and similarities between the $a_k$ parameters for each response choice as we did in Sections \ref{sec:rankplot} and \ref{sec:irtrank}.

\section{Summary and Future Directions}
The 2PL-NRM nested logit IRT model may be used to rank incorrect responses to FMCE questions by considering the calculated $a_k$ value as a measure of the correlation between choosing a particular response and the value of the $\theta$ parameter representing overall understanding of Newtonian mechanics (as measured by the FMCE).  We have shown that using random samples of a large data set can generate distributions of values for each $a_k$ that allow us to determine whether or not these parameters are meaningfully different, and we used Hedges' $g$ as a statistical measure to quantify these differences. We made particular choices regarding the values of $g$ that we consider to represent parameters that are approximately equal, those that are definitely different, and those that could go either way, and we have reported the value of $g$ for all comparisons to allow the reader to evaluate the validity of our claims or determine a ranking based on other choices of thresholds.

In many cases the responses that could not be determined to be definitely different or approximately the same as others are those that are rarely chosen by students. Future research will focus on clarifying these comparisons and trying to determine a robust ranking for all responses to every question. One way to accomplish this will be to perform similar analyses on other large data sets. Online data collection and analysis tools such as PhysPort's Data Explorer \cite{physportde} and the Learning About STEM Student Outcomes (LASSO) \cite{lasso} make this task much more achievable than it would been even a decade ago.

The results presented here are based on the assumption that students who understand more about physics will answer more questions correctly on the FMCE, and will also select better incorrect responses than students who understand less overall physics. We also ignored whether data were collected before or after instruction: we didn't care how students obtained their understanding, just that they had some level of understanding when they chose their responses. In future work, we will explore the implications of different assumptions for what makes one response better than another. 

One such assumption is that students will choose better responses after instruction than before instruction. This assumption is supported by the fact that (on average) students are more likely to choose correct responses after instruction than before: even low class-averaged gains from traditional instruction tend to be positive  \cite{Hake1998,Thornton2009,VonKorff2016}. A method consistent with this assumption would be to look for asymmetric transitions between response choices in matched pre/posttest data using a McNemar-Bowker chi-square test \cite{McNemar1947,Bowker1948}.

Another assumption that could be made is that students are more likely to choose correct responses after instruction if they chose better incorrect responses before instruction. Using conditional probabilities would allow us to identify a progression of responses to each question, with students moving up the progression being considered getting closer to correct. This is consistent with Thornton's conceptual dynamics in which students move between various views as they progress toward the correct response \cite{Thornton1997}.

Each of these methods could be used to test the rankings presented above and help clarify the ambiguously ranked responses. These methods and assumptions may also be applied to other research-based assessments to determine a robust ranking of incorrect responses for any multiple-choice question.

\begin{acknowledgements}
	We thank Sam McKagan and Ellie Sayre for providing access to data from PhysPort's Data Explorer and all of the instructors who were willing to share their students' FMCE responses. We also thank Kerry Gray, Nicholas Wright, Ian Griffin, and Ryan Moyer for their previous contributions as members of the research team. This project was supported by the National Science Foundation through grant DUE-1836470. 
\end{acknowledgements}

\appendix
\section{Additional 2PL-NRM Plots}
\begin{figure*}
    \centering
    \begin{tabular}{c@{}c@{}c@{}c}
    \includegraphics[width = 0.25 \textwidth]{irtQ1.pdf}&
    \includegraphics[width = 0.25 \textwidth]{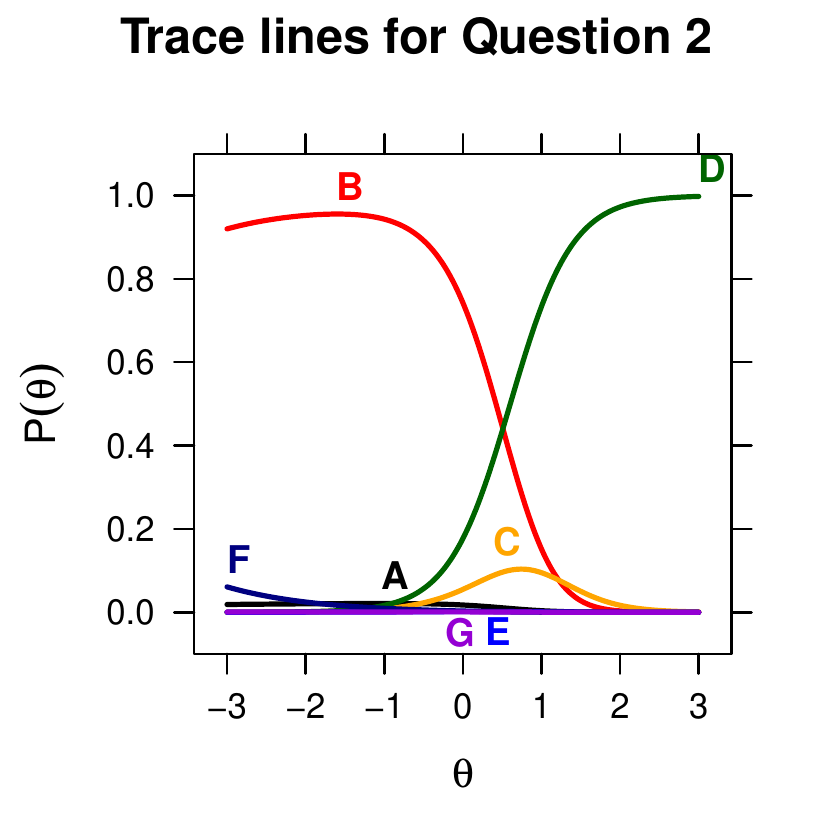}&
    \includegraphics[width = 0.25 \textwidth]{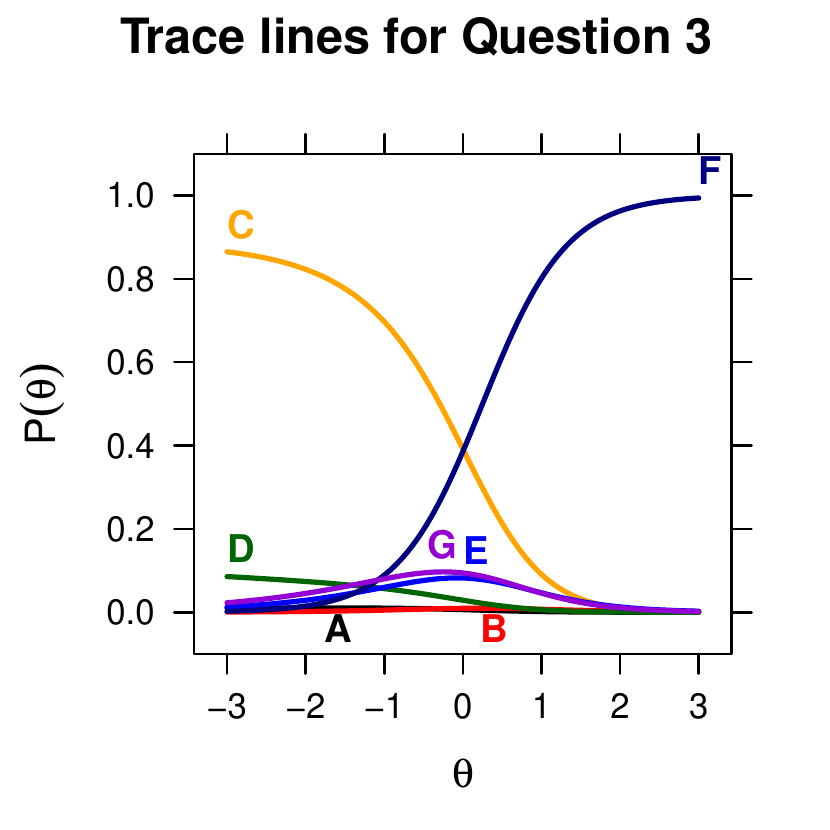}&
    \includegraphics[width = 0.25 \textwidth]{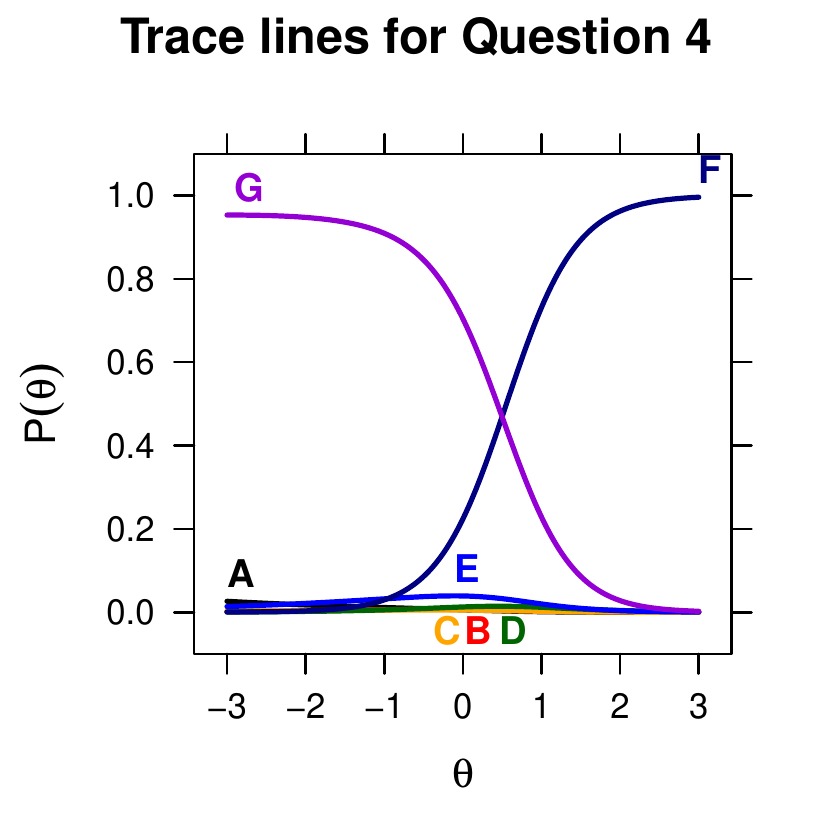}\\
    \includegraphics[width = 0.25 \textwidth]{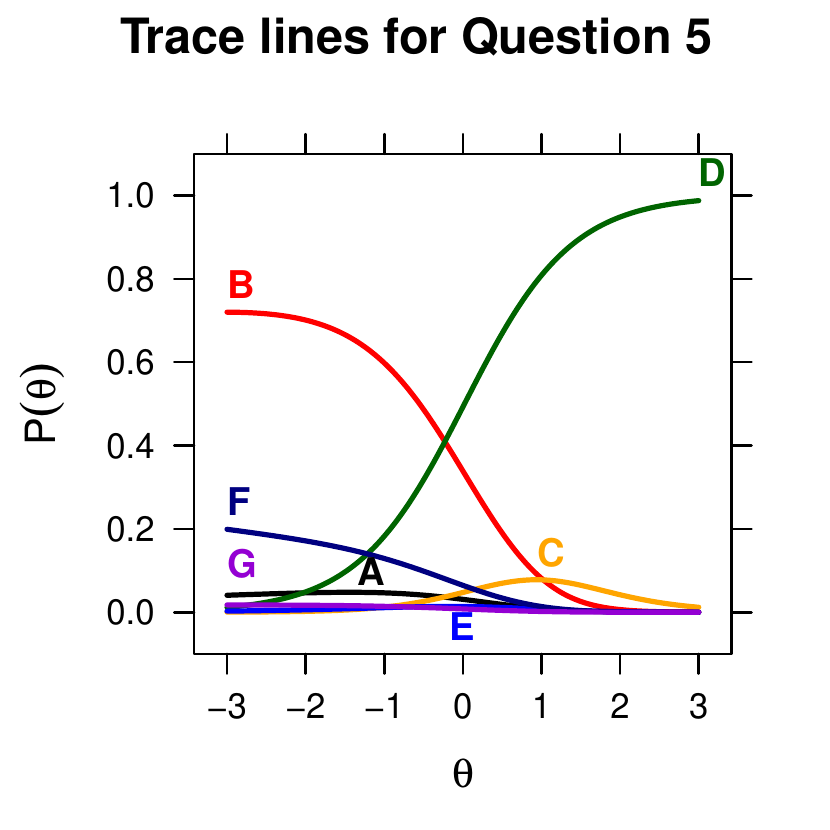}&
    \includegraphics[width = 0.25 \textwidth]{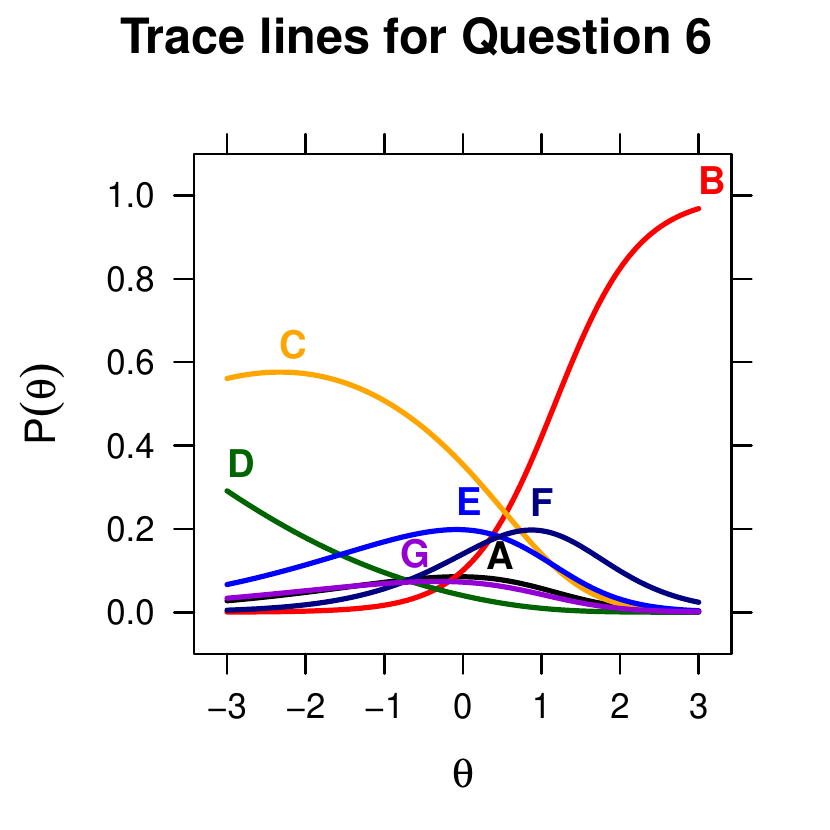}&   
    \includegraphics[width = 0.25 \textwidth]{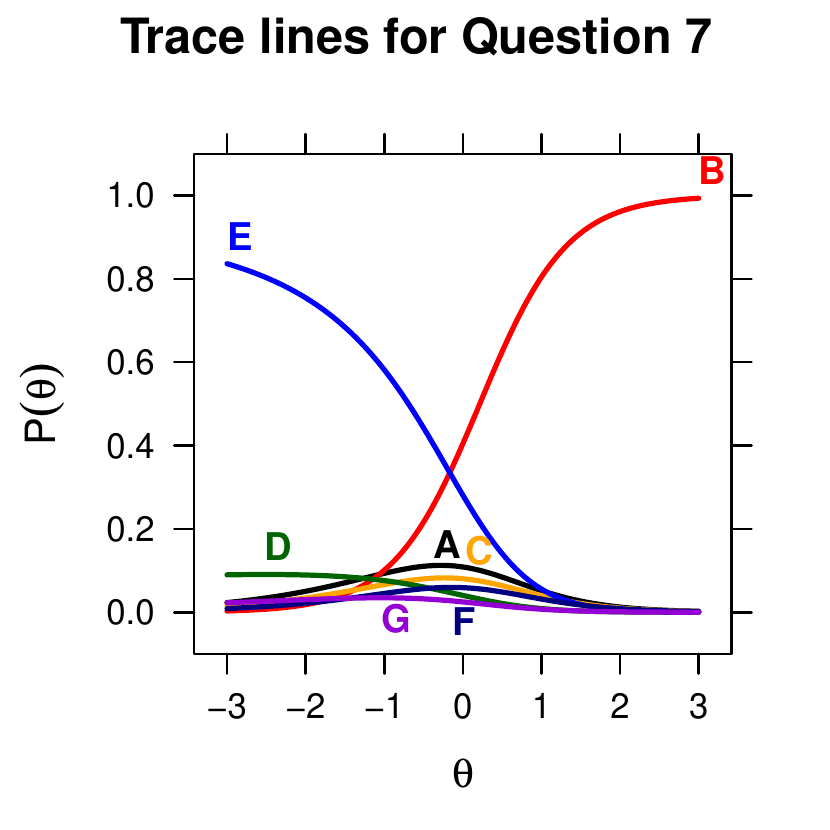}&
    \includegraphics[width = 0.25 \textwidth]{irtQ8.pdf}\\
    \includegraphics[width = 0.25 \textwidth]{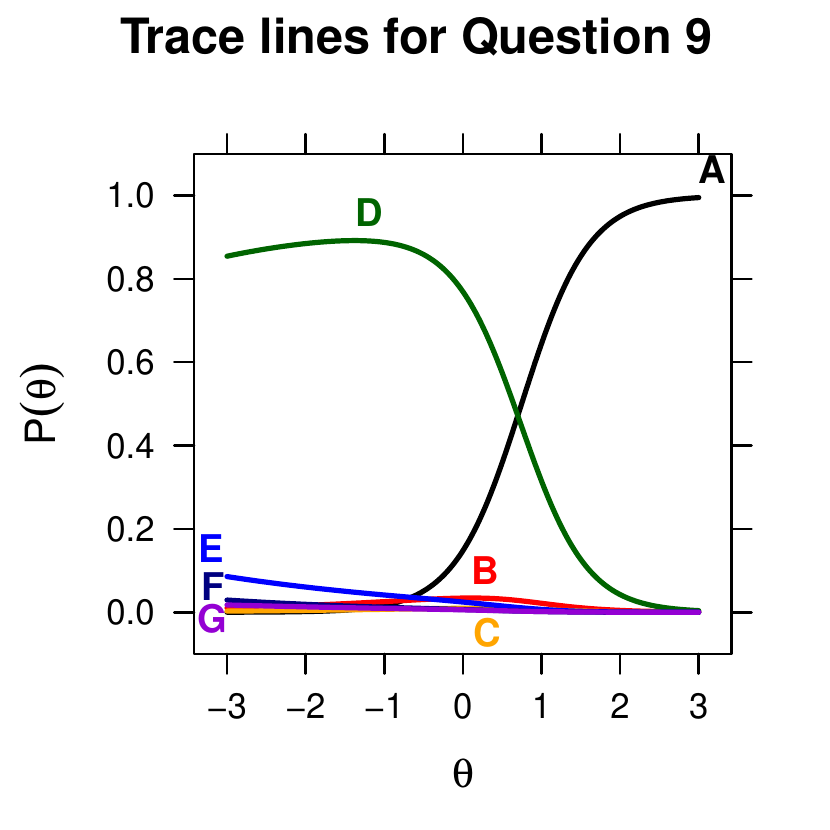}&        \includegraphics[width = 0.25 \textwidth]{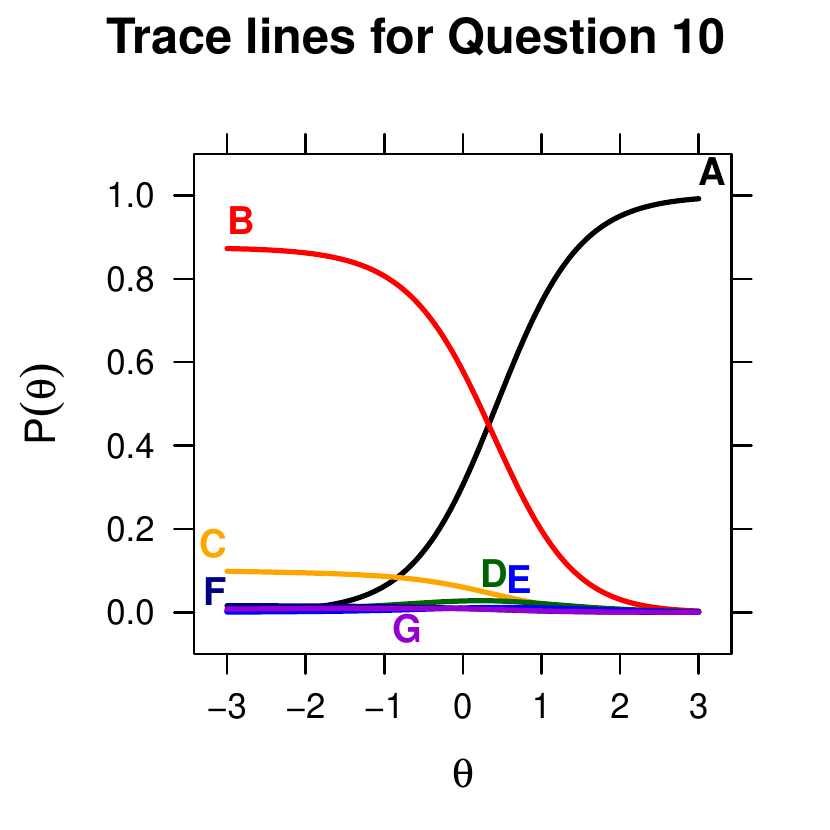}&        \includegraphics[width = 0.25 \textwidth]{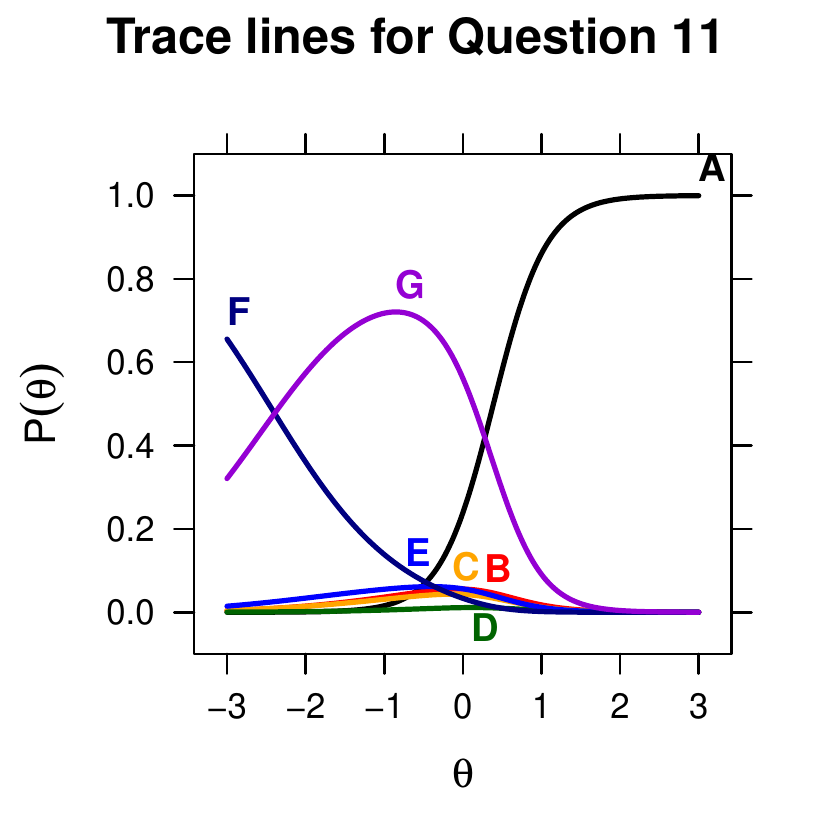}&        \includegraphics[width = 0.25 \textwidth]{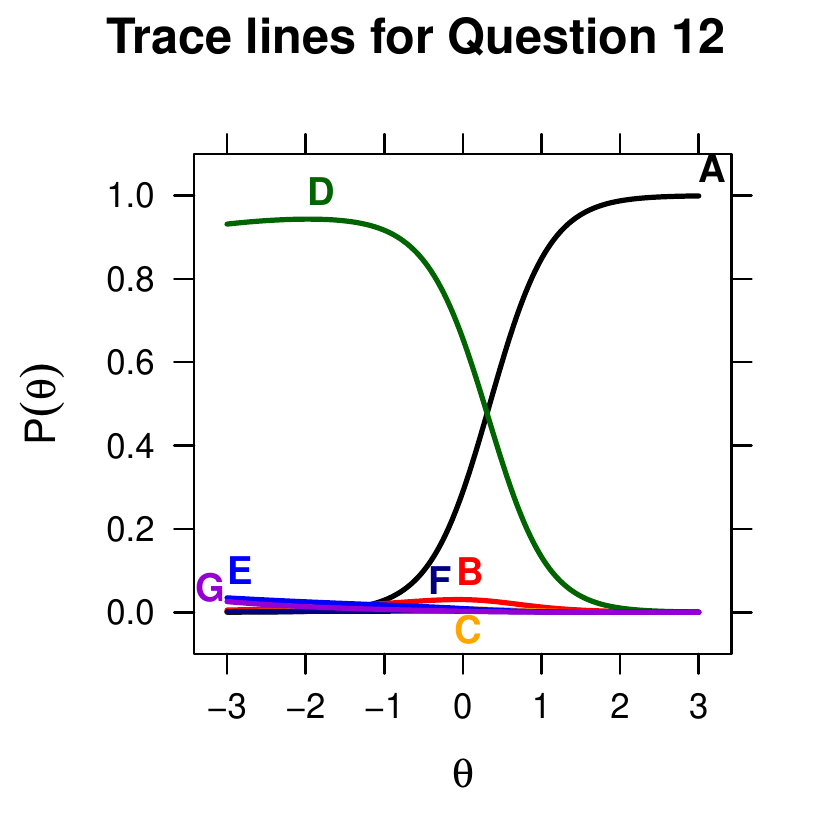}\\        \includegraphics[width = 0.25 \textwidth]{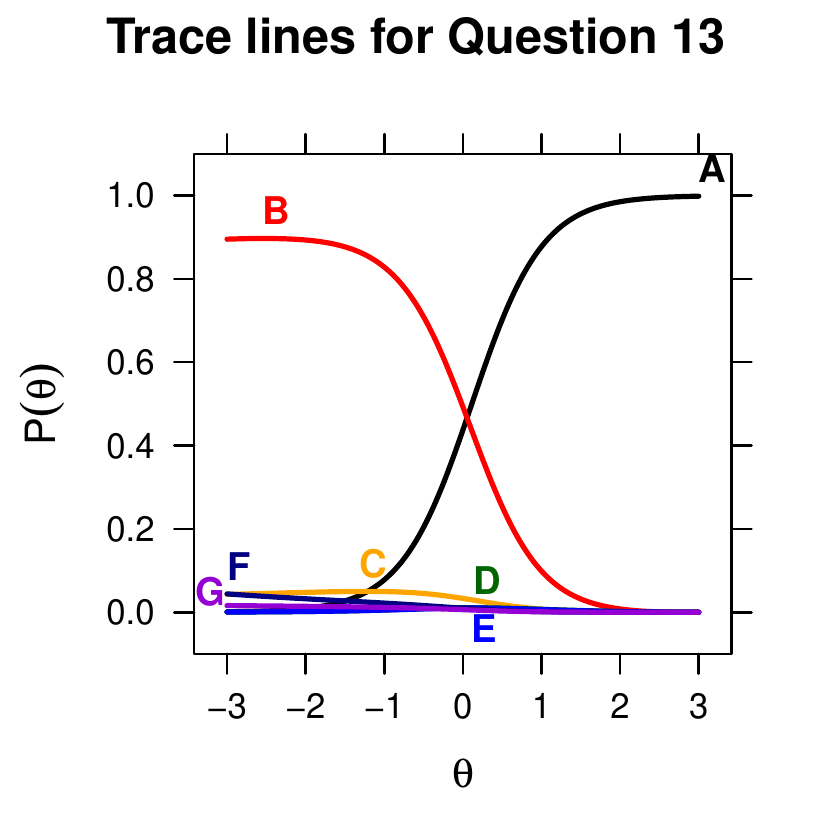}&        \includegraphics[width = 0.25 \textwidth]{irtQ14.pdf}&        \includegraphics[width = 0.25 \textwidth]{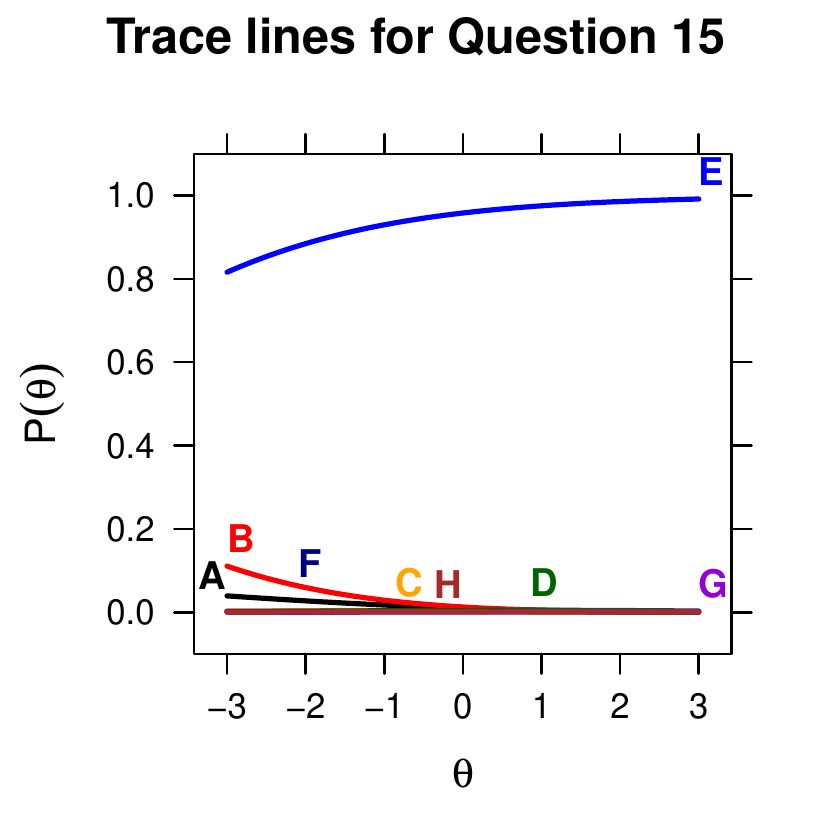}&        \includegraphics[width = 0.25 \textwidth]{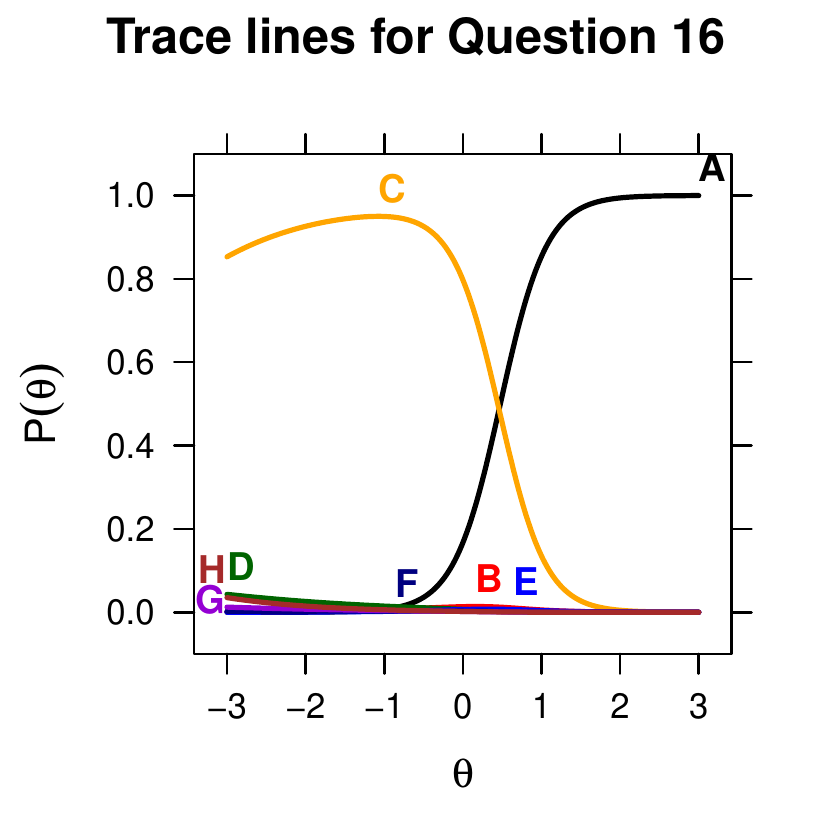}\\        \includegraphics[width = 0.25 \textwidth]{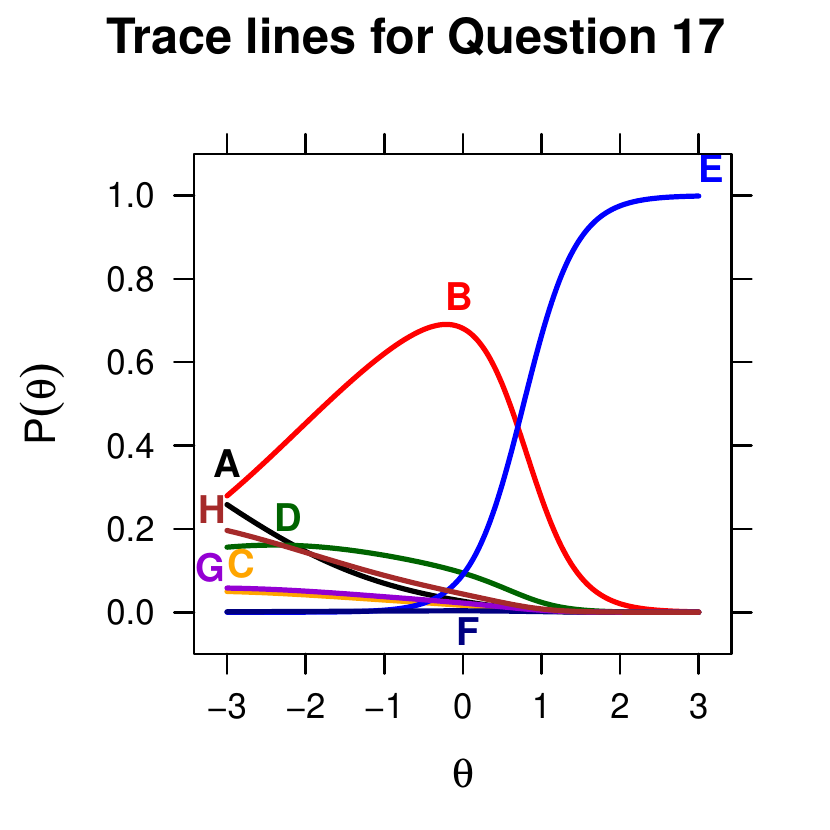}&        \includegraphics[width = 0.25 \textwidth]{irtQ18.pdf}&        \includegraphics[width = 0.25 \textwidth]{irtQ19.pdf}&        \includegraphics[width = 0.25 \textwidth]{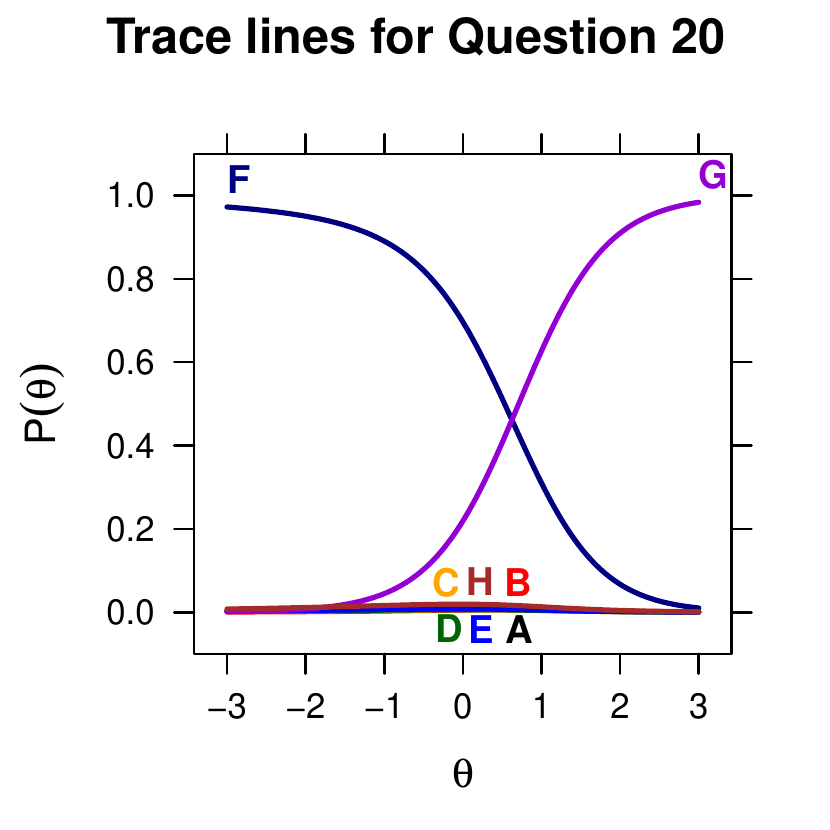}\\        \end{tabular}
    \caption{IRT plots from the 2PL-NRM nested logit model for questions 1--20:  probability of choosing each response as a function of the person parameter $\theta$ (representing overall understanding of Newtonian mechanics). Plots created using the mirt and directlabels packages in R \cite{mirt,directlabels,r}.}
    \label{fig:2pl-nrm120}
\end{figure*}

\begin{figure*}
    \centering
    \begin{tabular}{c@{}c@{}c@{}c}
    \includegraphics[width = 0.25 \textwidth]{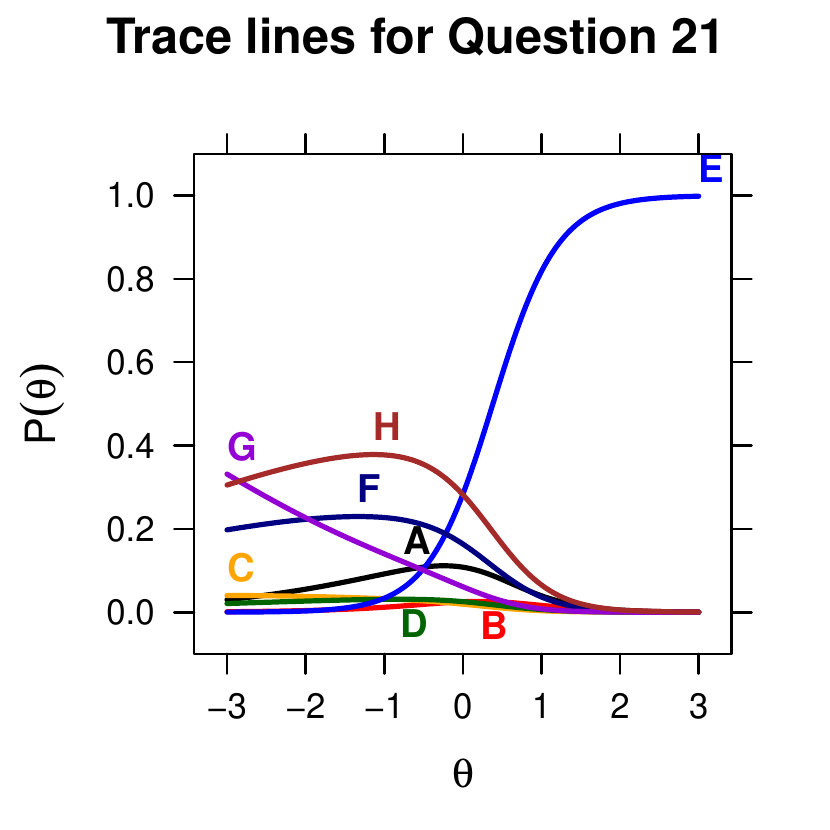}&        \includegraphics[width = 0.25 \textwidth]{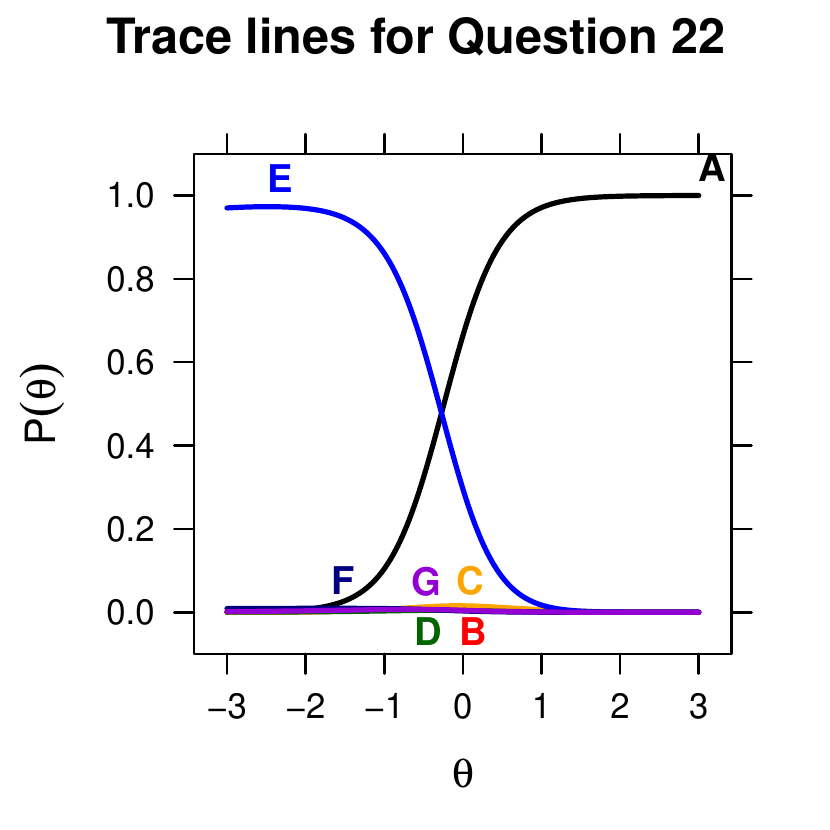}&        \includegraphics[width = 0.25 \textwidth]{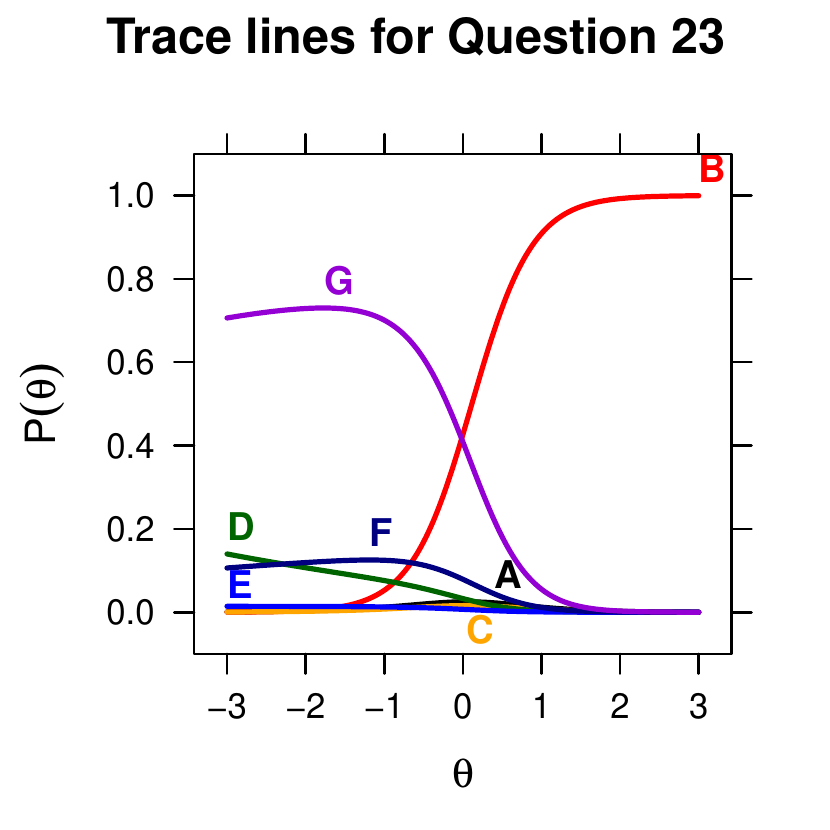}&        \includegraphics[width = 0.25 \textwidth]{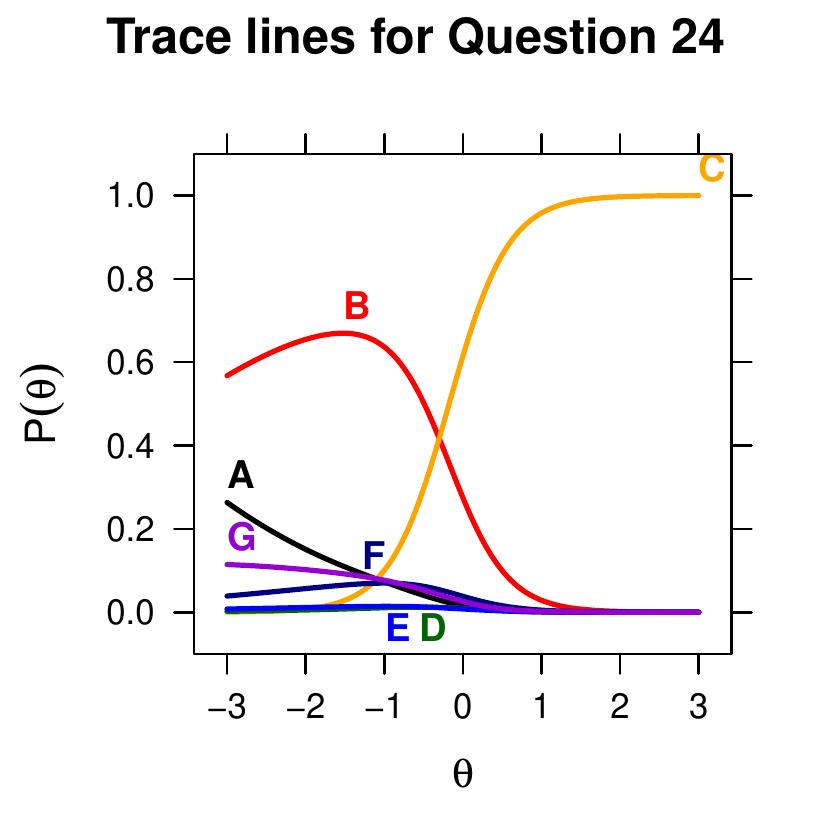}\\        \includegraphics[width = 0.25 \textwidth]{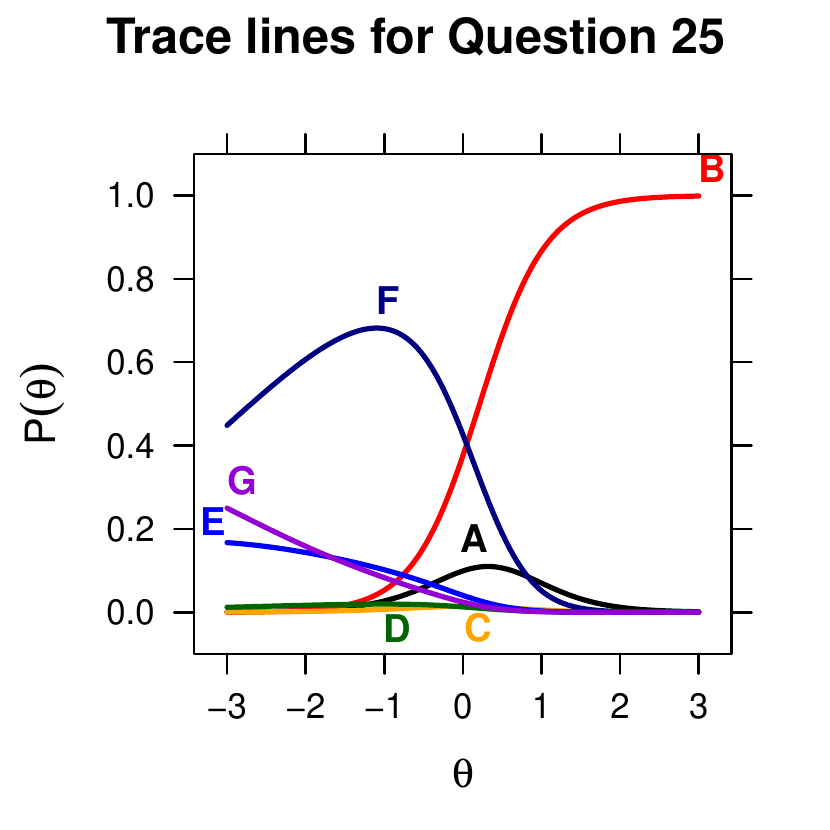}&        \includegraphics[width = 0.25 \textwidth]{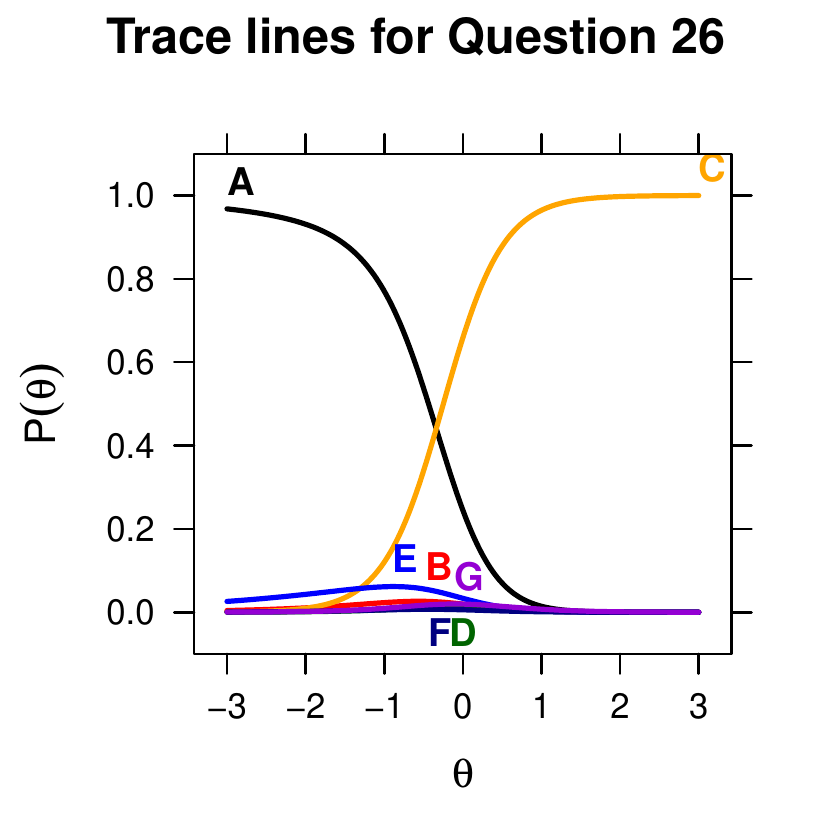}&        \includegraphics[width = 0.25 \textwidth]{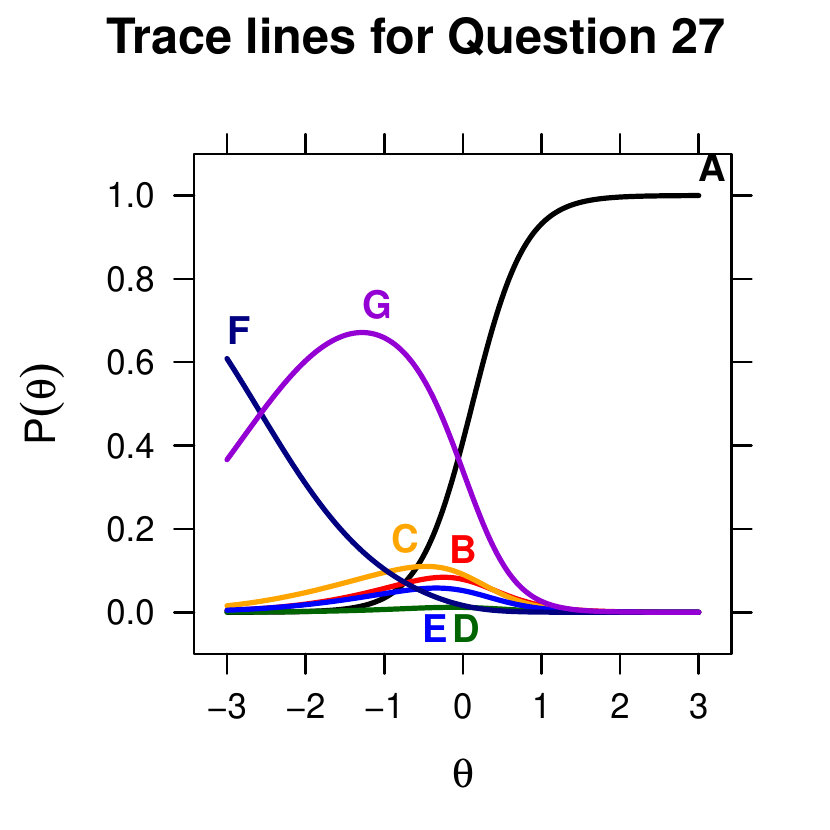}&        \includegraphics[width = 0.25 \textwidth]{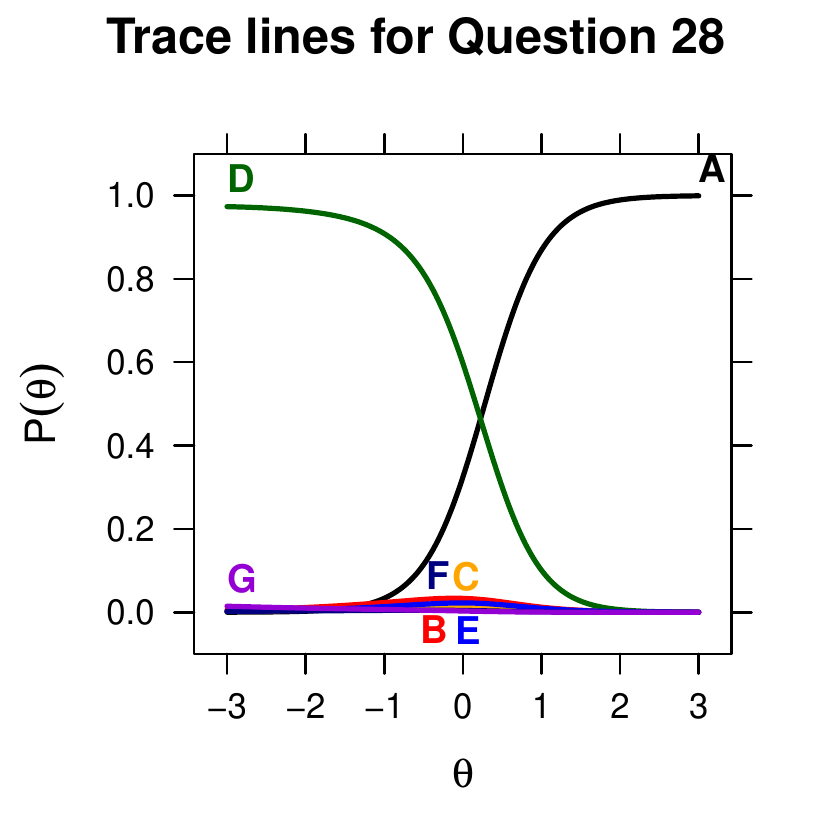}\\        \includegraphics[width = 0.25 \textwidth]{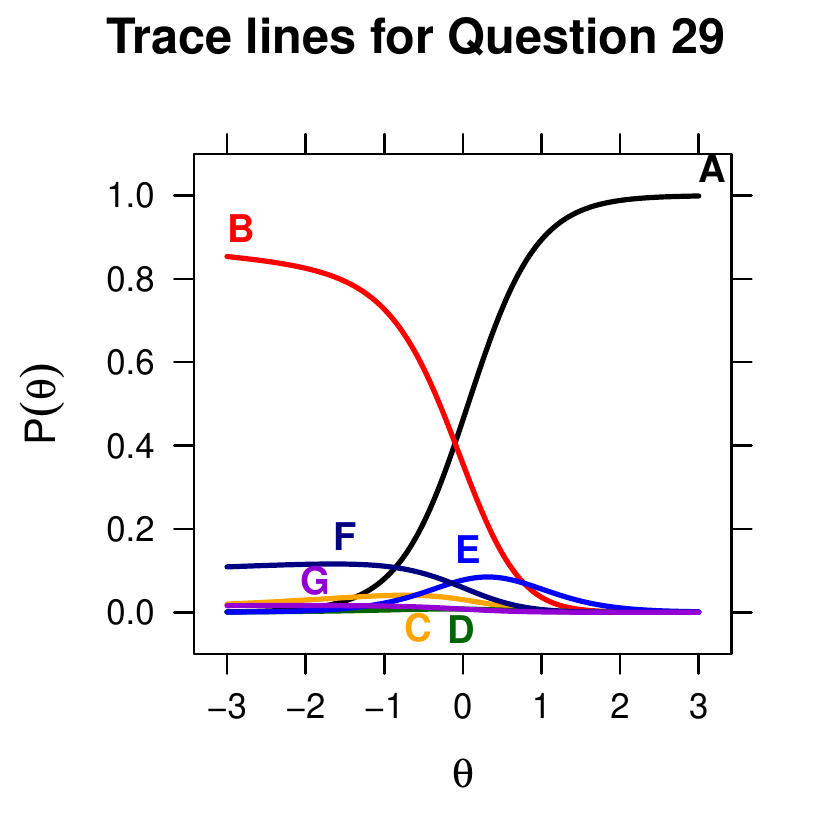}&        \includegraphics[width = 0.25 \textwidth]{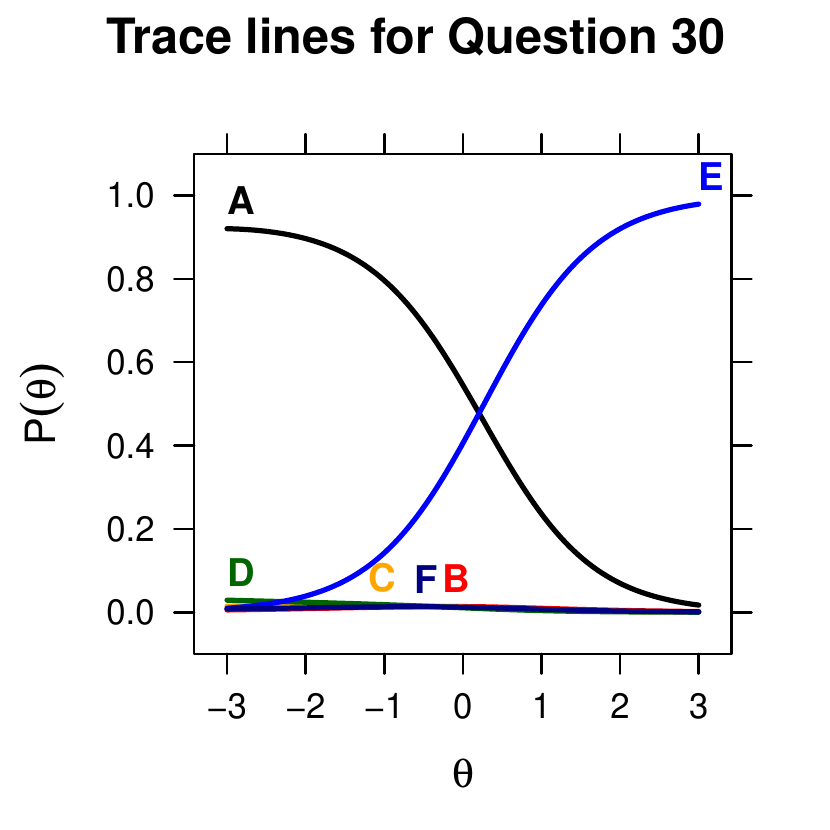}&        \includegraphics[width = 0.25 \textwidth]{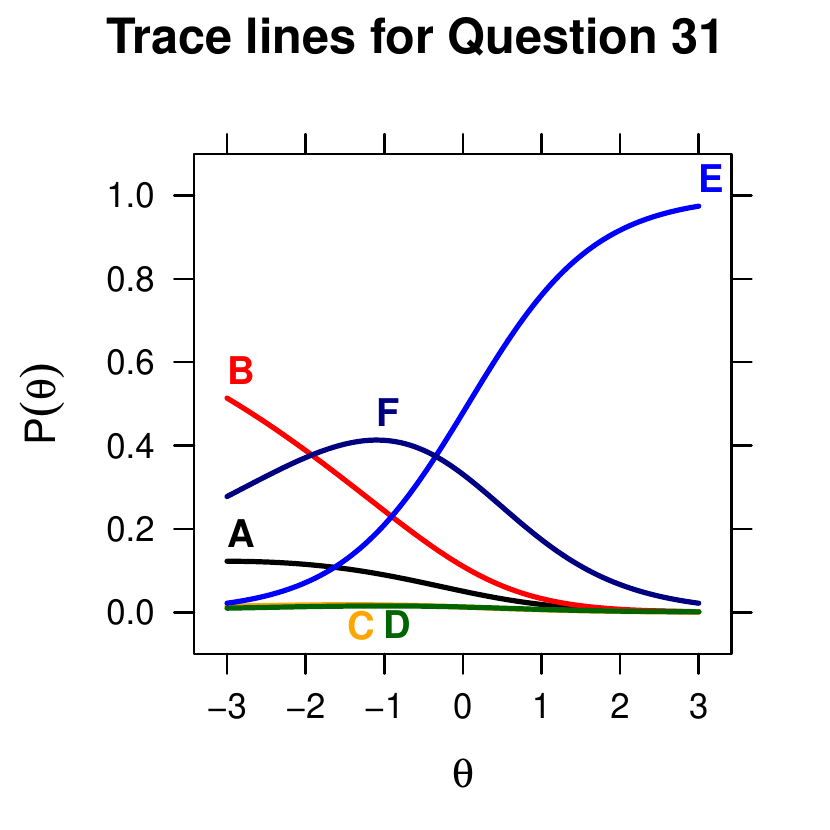}&        \includegraphics[width = 0.25 \textwidth]{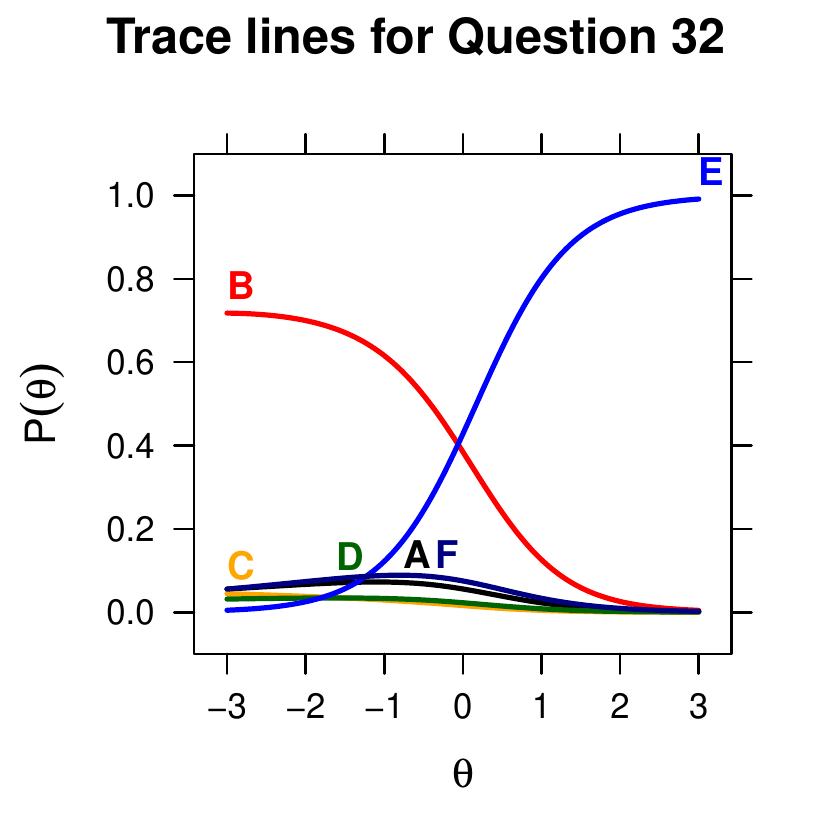}\\        \includegraphics[width = 0.25 \textwidth]{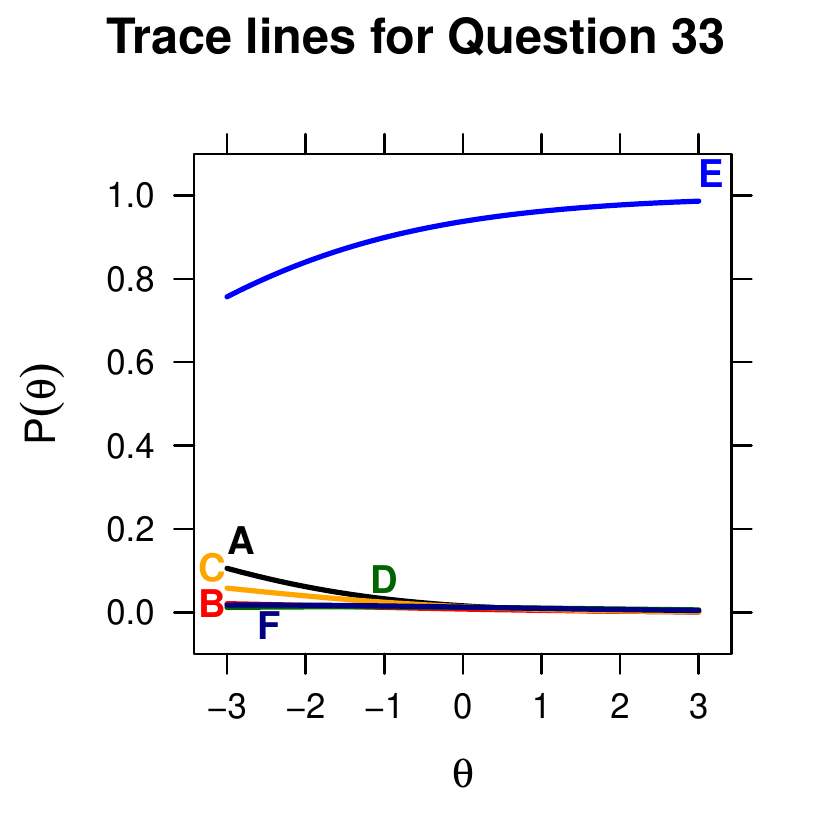}&        \includegraphics[width = 0.25 \textwidth]{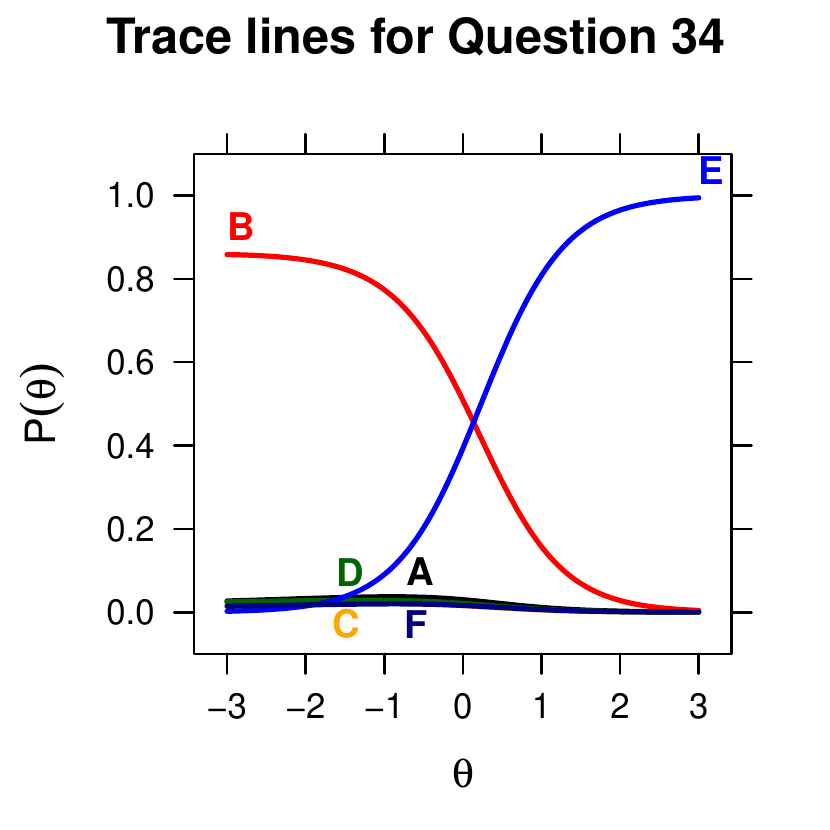}&        \includegraphics[width = 0.25 \textwidth]{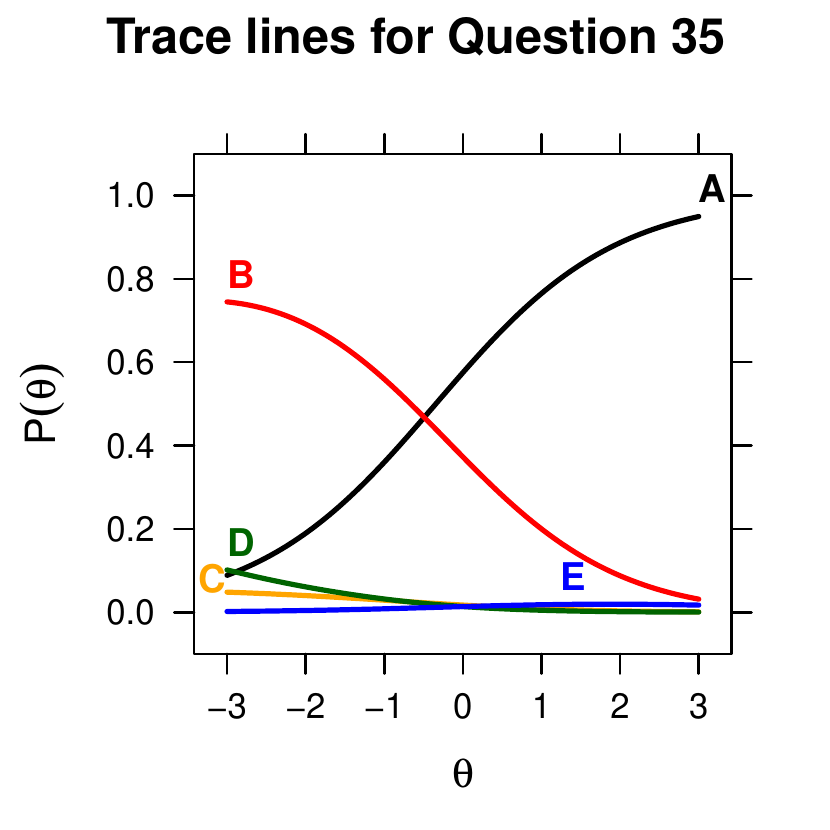}&        \includegraphics[width = 0.25 \textwidth]{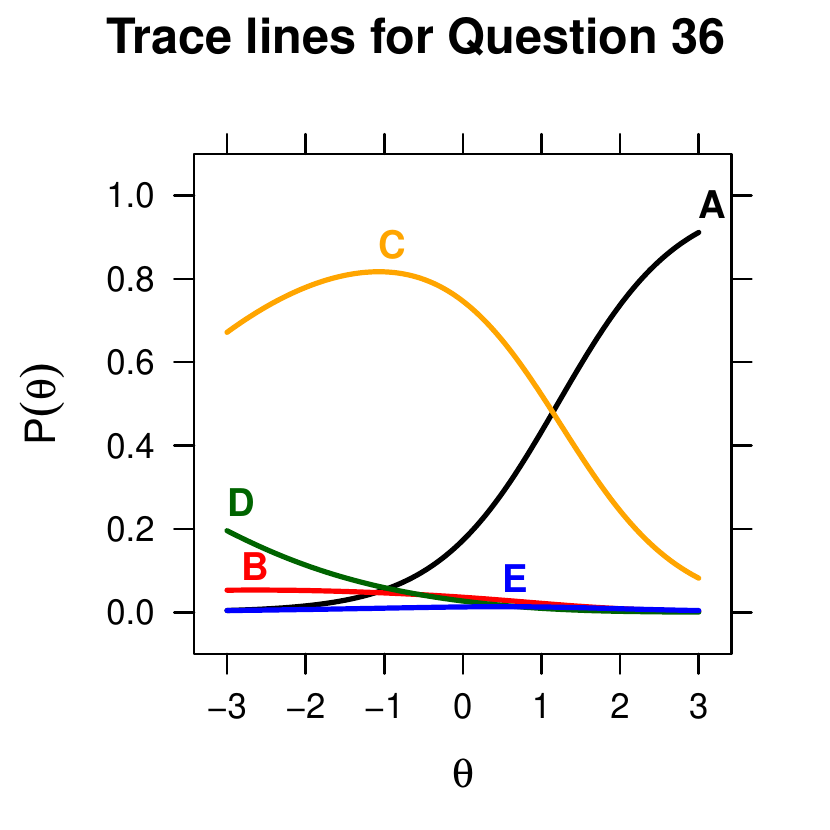}\\        \includegraphics[width = 0.25 \textwidth]{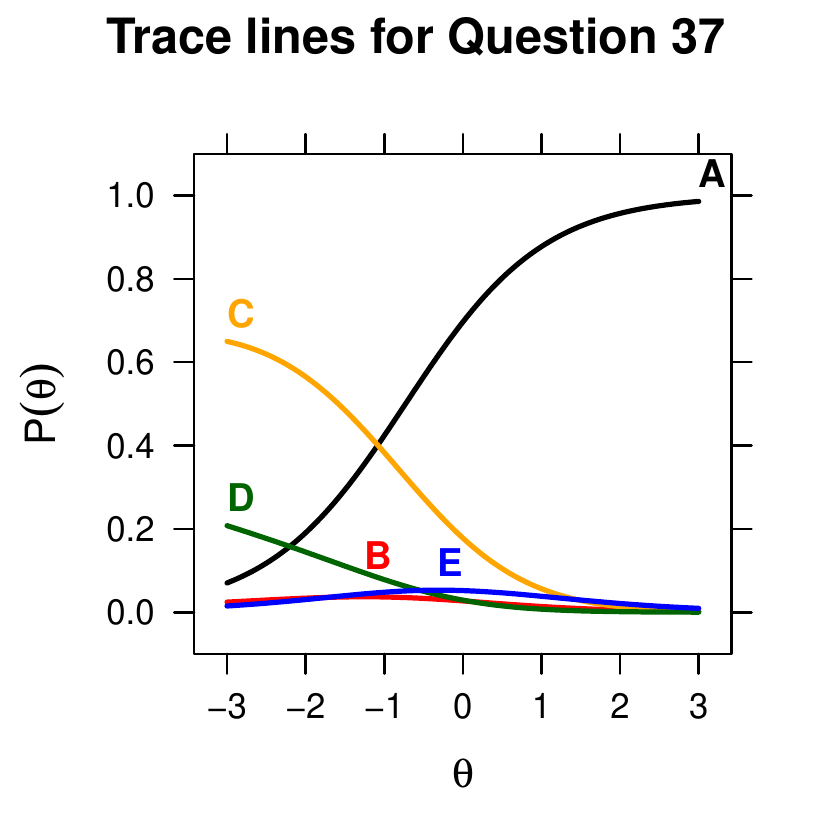}&        \includegraphics[width = 0.25 \textwidth]{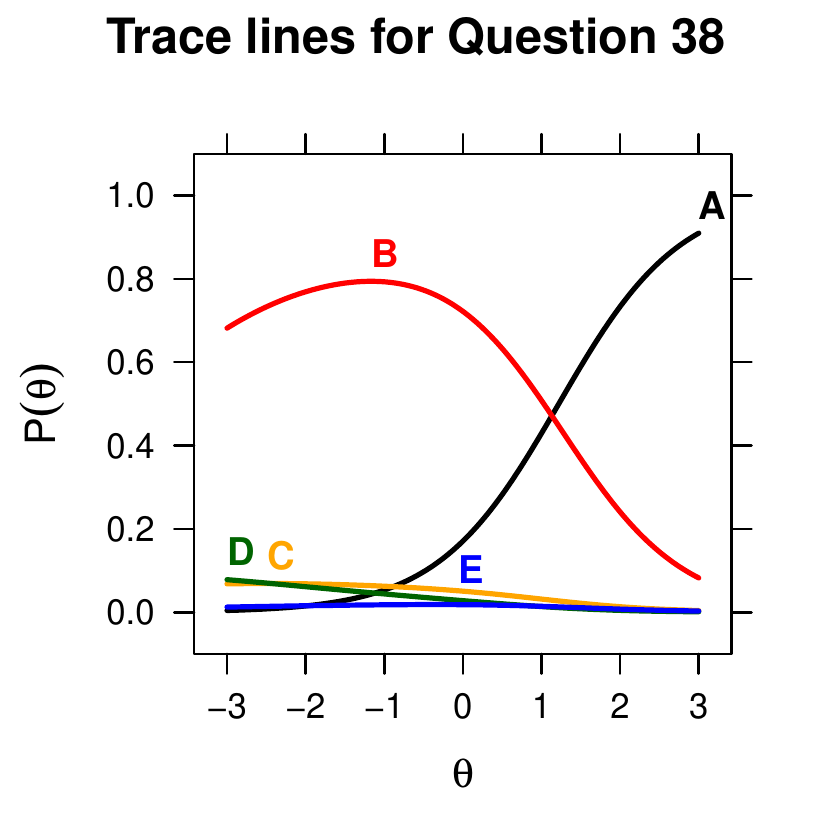}&        \includegraphics[width = 0.25 \textwidth]{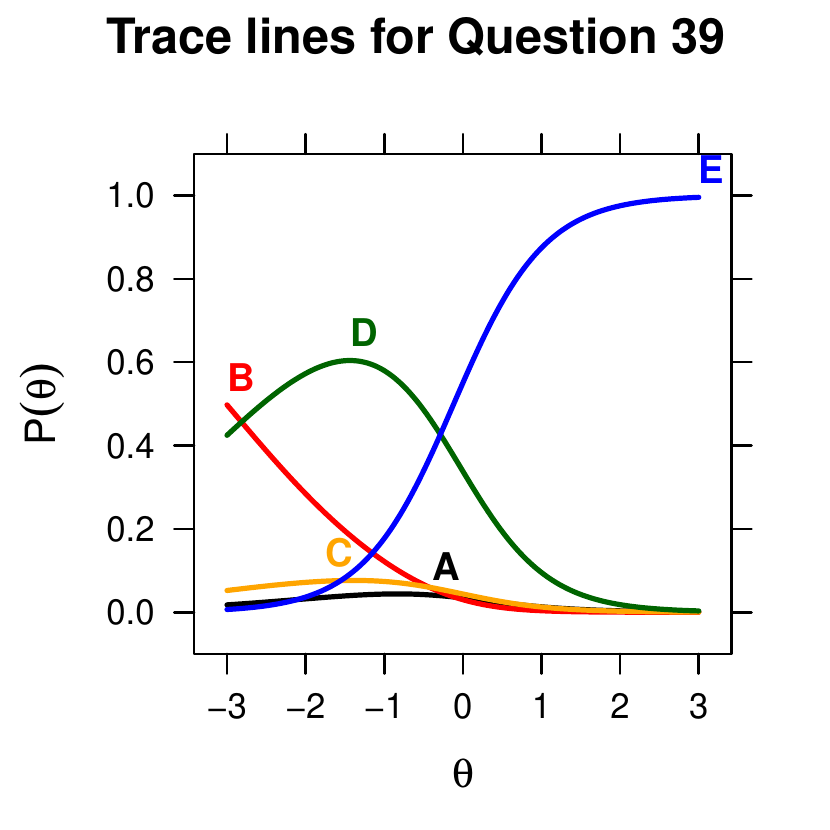}&        \includegraphics[width = 0.25 \textwidth]{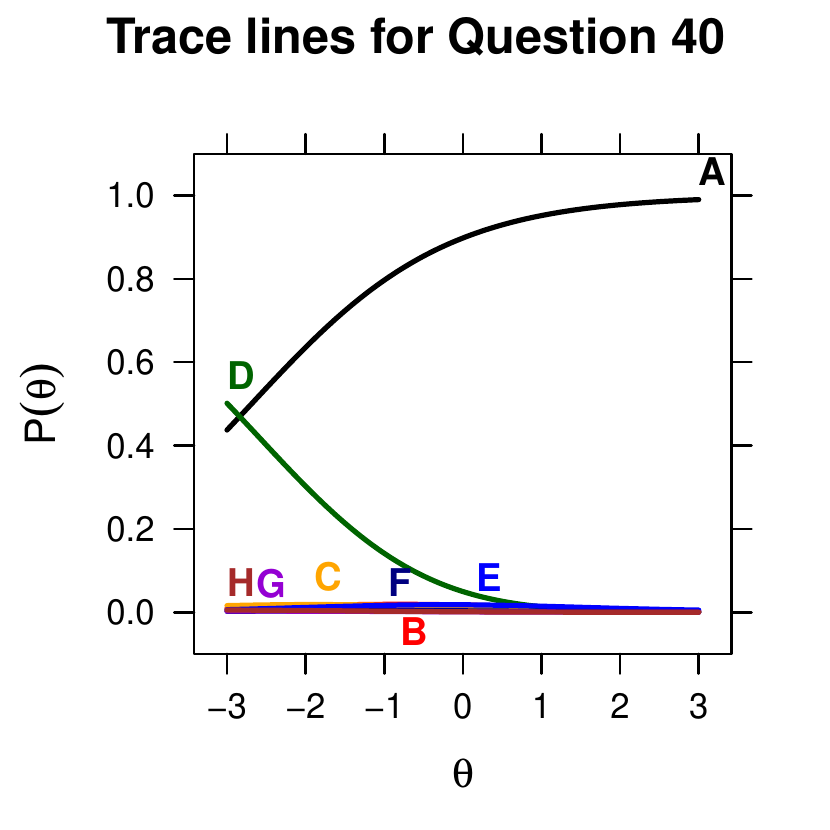}\\    \end{tabular}
    \caption{IRT plots from the 2PL-NRM nested logit model for questions 21--40.}
    \label{fig:2pl-nrm2140}
\end{figure*}

\begin{figure*}
    \centering
    \begin{tabular}{c@{}c@{}c@{}c}
    \includegraphics[width = 0.25 \textwidth]{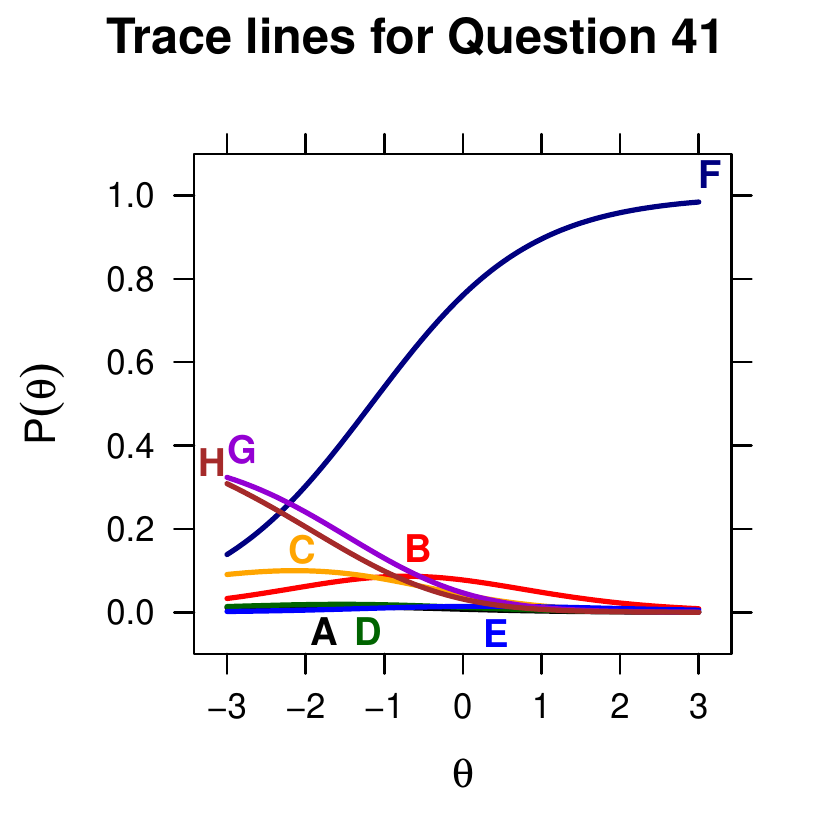}&        \includegraphics[width = 0.25 \textwidth]{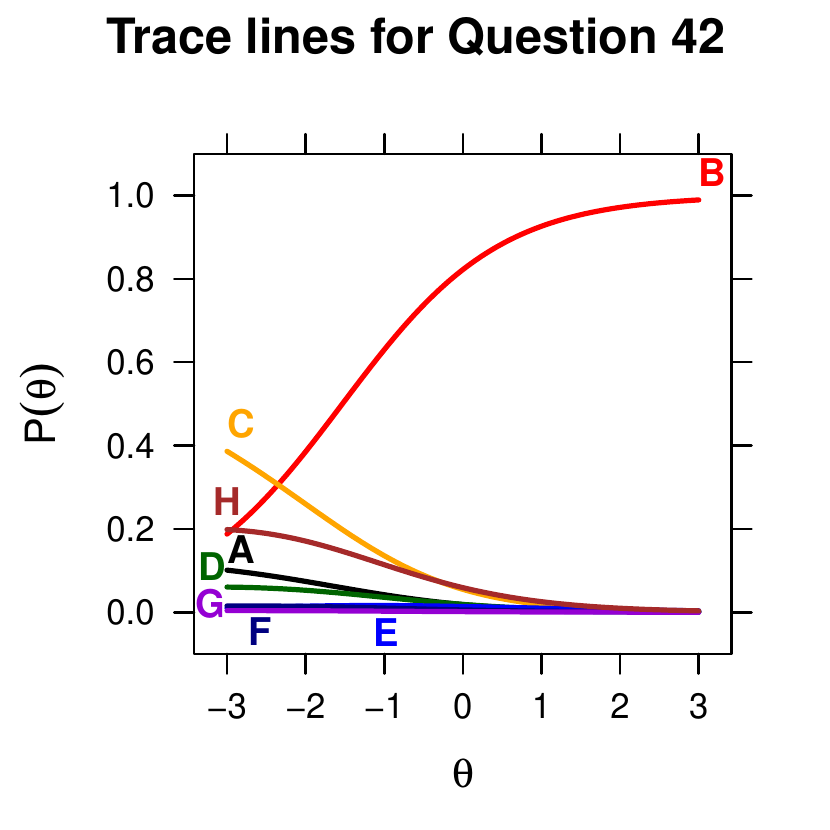}&        \includegraphics[width = 0.25 \textwidth]{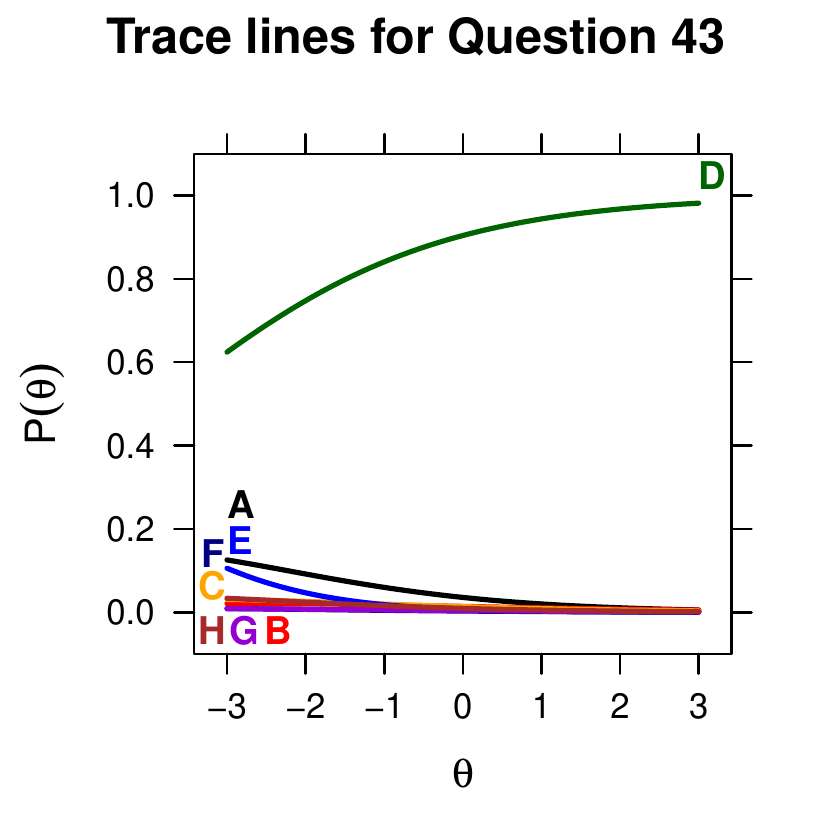}&        \includegraphics[width = 0.25 \textwidth]{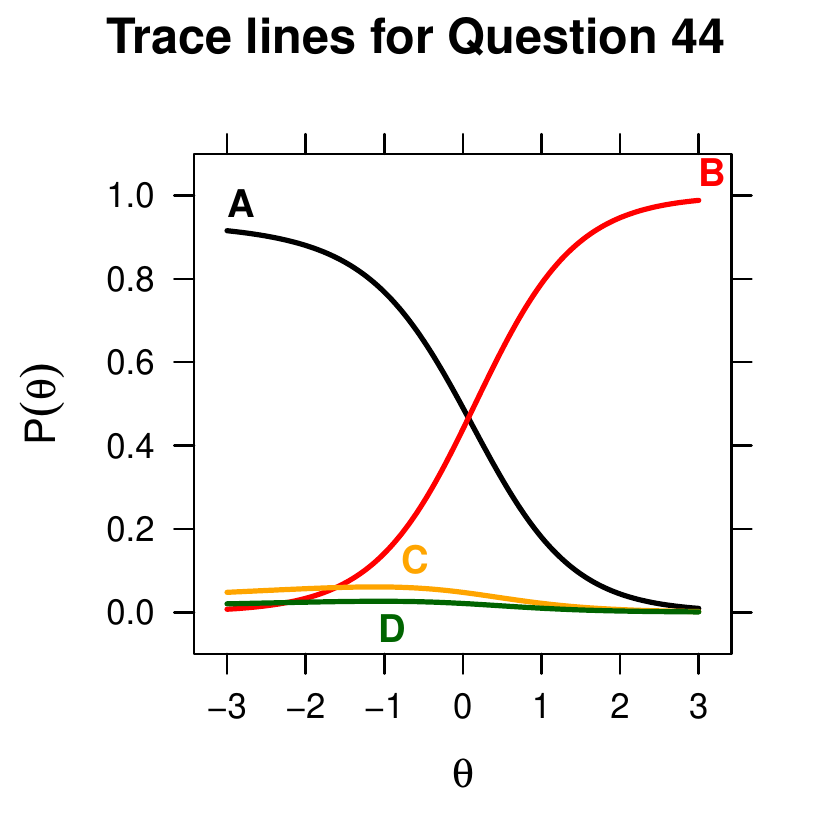}\\        \includegraphics[width = 0.25 \textwidth]{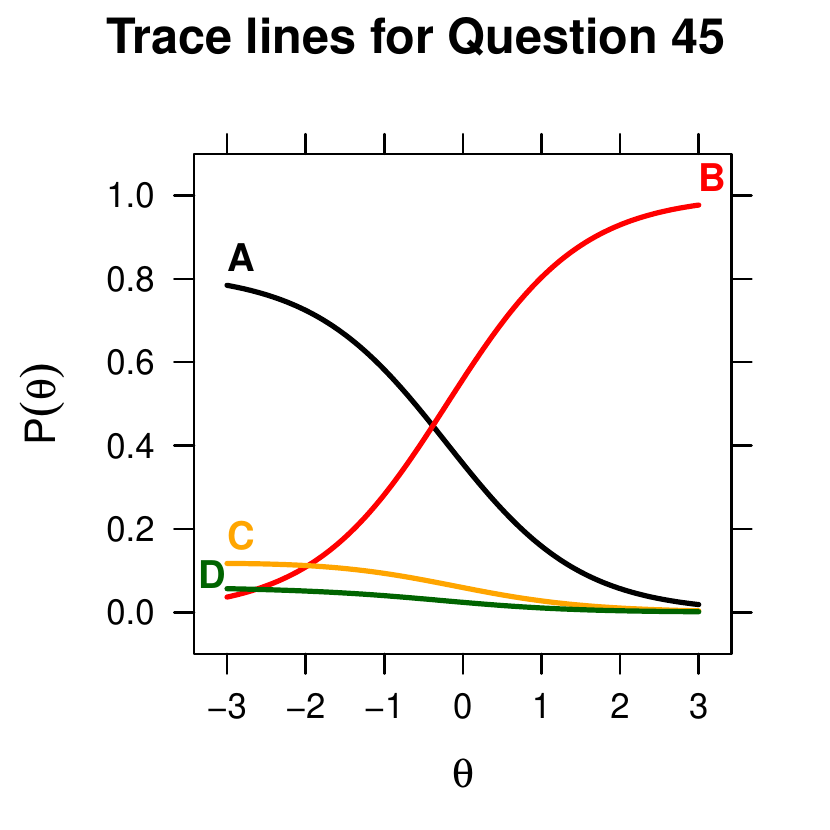}&        \includegraphics[width = 0.25 \textwidth]{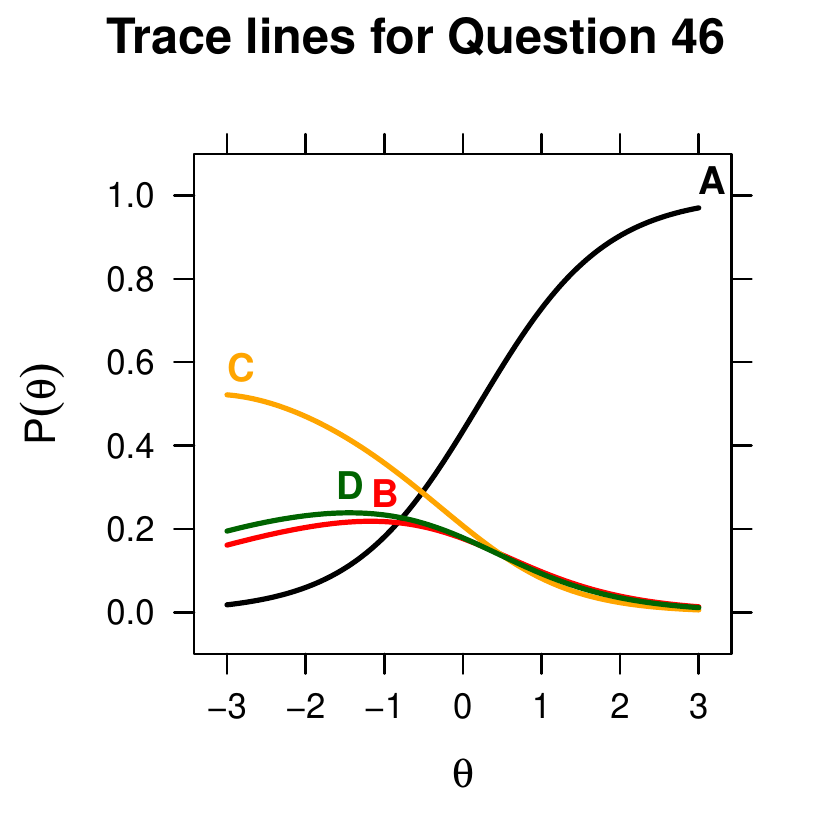}&        \includegraphics[width = 0.25 \textwidth]{irtQ47.pdf}&\\    \end{tabular}
    \caption{IRT plots from the 2PL-NRM nested logit model for questions 41--47.}
    \label{fig:2pl-nrm4147}
\end{figure*}

Figures \ref{fig:2pl-nrm120}--\ref{fig:2pl-nrm4147} are included to allow the reader to see how the rankings in Tables \ref{rankTable126} and \ref{rankTable2747} relate to the probability of choosing each response, given the value of a student's $\theta$ parameter. Each curve is labeled near the maximum value (with slight adjustments to avoid overlapping labels and curves), so the horizontal location of the label provides an approximate ranking of the responses (right is better, left is worse). Ambiguously ranked responses often show up as lines near zero probability for all values of $\theta$.

\bibliography{TIS.bib}

\end{document}